\documentclass{article}

\usepackage{geometry,amsmath,amssymb,graphicx, comment, authblk, url, hyperref}

\usepackage{float}
\usepackage[caption = false]{subfig}

\usepackage{pgfplots}

\usepackage{tikz}

\tikzstyle{medium_block} = [rectangle, draw, text width=8em, text centered, rounded corners, minimum height=3em]
\tikzstyle{mb_block} = [rectangle, draw, text width=10em, text centered, rounded corners, minimum height=3em]
\tikzstyle{block} = [rectangle, draw, text width=6em, text centered, rounded corners, minimum height=3em]
\tikzstyle{big_block} = [rectangle, draw, text width=15em, text centered, rounded corners, minimum height=3em]

\newcommand{\numberfield}[1]{\ensuremath{\mathbb{#1}}} 

\title{Principle of Information Increase: An Operational Perspective of Information Gain in the Foundations of Quantum Theory}
\author{Yang Yu}	
\author{Philip Goyal}	
\affil{University at Albany~(SUNY), NY, USA}
\date{}

\begin{document}
	
	\maketitle
	
	\begin{abstract}
		A measurement performed on a quantum system is an act of gaining information about its state, a view that is widespread in practical and foundational work in quantum theory. However, the concept of information in quantum theory reconstructions is multiply-defined, and its conceptual foundations remain surprisingly under-explored. In this paper, we investigate the gain of information in quantum measurements from an operational viewpoint. We show that the continuous extension of the Shannon entropy naturally admits \emph{two} distinct measures of information gain, \emph{differential information gain} and \emph{relative information gain}, and that these have radically different characteristics. In particular, while differential information gain can increase or decrease as additional data is acquired, relative information gain consistently grows, and moreover exhibits asymptotic indifference to the data or choice of Bayesian prior.  
		
		In order to make a principled choice between these measures, we articulate a \emph{Principle of Information Increase}, which incorporates Summhammer's proposal  \cite{Summhammer1994, Summhammer1999} that more data from measurements leads to more knowledge about the system, and also takes into consideration black swan events. This principle favors differential information gain as the more relevant metric in two-outcome quantum systems, and guides the selection of priors for these information measures. Finally, we show that, of the beta distribution priors, the Jeffreys' binomial prior is the prior ensures maximal \emph{robustness} of information gain to the particular data sequence obtained in a run of experiments.

	\end{abstract}
	
	\section{Introduction}
	\label{sec:intro}
	
	A measurement performed on a quantum system is an act of acquiring information about its state. This informational perspective on quantum measurement is widely embraced in practical applications such as quantum tomography~\cite{Patra_2007,Madhok2014,Quek2021,Gupta2021}, Bayesian experimental design~\cite{McMichael2021}, and informational analysis of experimental data~\cite{Placek2017,MA2018}.  It is also embraced in foundational research.
	In particular, information assumes a central role in the quantum reconstruction program~\cite{Grinbaum03,Goyal2022c}, which seeks to elucidate the fundamental physical origins of quantum theory by deriving its formalism from information-inspired postulates~\cite{Brukner2009,Goyal2008,GKS-PRA,Goyal2014,Caticha2011,Masanes2013,Raedt2016,Hohn2017,Aravinda2017,Czekaj2017,Chiribella2018}. 
	Nonetheless, in the foundational exploration of quantum theory, the concept of information is articulated and formalized in many different ways, which raises the question of whether there exists a more systematic basis for choosing how to formalize the concept of information within this domain.
	In this paper, we scrutinize the notion of information from an operational standpoint, and propose a physically intuitive postulate to determine the appropriate information gained from measurements.
	
	In both tomographic applications and reconstruction of quantum theory, the focus often lies on probability distributions of physical parameters or quantities, which are updated based on the measurement results.  For example, the probability distribution $\Pr(x|D)$ of a quantity~$x$, which is updated from a prior probability distribution given the results, $D$, obtained from a series of measurements performed on identical copies of a system.  It is natural to consider using Shannon entropy to quantify the information gained from this updated distribution. However, Shannon entropy is limited to discrete distributions, whereas physical quantities and their associated probability distributions can be continuous.
	
	The question thus arises:~what is a suitable measure for quantifying the information obtained from real data, especially for quantities associated with continuous probability distributions? 
	One potential solution is to employ Kullback-Leibler~(KL) divergence, also known as the relative entropy, $H(x|D) = \int \Pr(x | D) \ln \frac{\Pr(x|D)}{\Pr(x|I)} dx$, where~$\Pr(x|I)$ represents the prior distribution of~$x$, and~$\Pr(x|D)$ represents the posterior distribution of~$x$ updated with the data~$D$. This quantity is commonly referred to as the information gain from the prior distribution to the posterior distribution, and is widely used.
	
	Since the KL divergence is non-negative and invariant under changes of coordinates, it appears to be a reasonable generalization of the Shannon entropy for continuous probability distributions.  However, there are situations where \emph{information gain} defined in terms of the KL divergence does not have a unique representation. Consider a scenario where one has acquired a series of data~$D$, and one proceeds to take \emph{additional} measurements, obtaining additional data~$D'$. What is the additional information gain pertaining to $D'$? Using the KL divergence, there are two distinct ways to express the information related to this additional data. The first, to which we refer henceforth as the~\textit{differential information gain}, is simply the difference between the information gain from the combined dataset~$\{D,D'\}$ and the information gain from~$D$ alone~(see Fig.~\ref{fig:differential-information-gain}).  The second, to which we refer as the~\textit{relative information gain}, is given by the KL divergence of the posterior distribution after obtaining the complete dataset $\{D,D'\}$ compared to the posterior distribution after receiving data $D$ alone~(see Fig.~\ref{fig:relative-information-gain}).
	These two measures of information gain exhibit notably different characteristics. For instance, whether the differential information gain increases or decreases when data~$D'$ is acquired depends on the choice of the prior distribution over the parameter, while the relative information gain consistently increases regardless of the choice of prior.

	As we shall discuss in Section~\ref{sec:background}, both of these measures can be viewed as arising as a consequence of seeking to generalize the Shannon entropy to continuous probability distributions.  In order to determine which of these options is most appropriate for our purposes, we seek a physically intuitive informational postulate to guide our selection. The first criterion comes from the intuitive notion proposed by Summhammer~\cite{Summhammer1994,Summhammer1999} that \emph{more data from measurements leads to more knowledge about the system}.  This idea has its origin in the observation that, as we conduct more measurements to determine the value of a physical quantity, the measurement uncertainty tends to decrease.  In the following, we employ information theory to formalize this idea, requiring that the information gain associated with additional data should be positive.  We find that relative information gain is consistently non-negative, whereas the positivity of differential information gain hinges on the choice of the prior distribution.
	
However, in contrast to Summhammer's criterion, we argue that, under certain circumstances, negative information gain due to acquisition of additional data~$D'$ is also meaningful.  Take, for instance, the occurrence of a \emph{black swan event}, an event so rare and unexpected that it significantly increases one's uncertainty about the color of swans.  If the gain of information is considered to result from a reduction in the degree of uncertainty, the information gain associated with the observation of a black swan should indeed be negative.  By combining this observation with Summhammer's criterion, we are lead to the \emph{Principle of Information Increase:} the information gain from additional data should be positive asymptotically, and negative in extreme cases.  On the basis of the Principle of Information Increase, we show that differential information gain is the more appropriate measure.  

In addition, we formulate a new criterion, \emph{the robustness of information gain}, for selecting priors to use with the differential information gain.  The essential idea behind this criterion is as follows.  If the result of the additional data~$D'$ is fixed, then the information gain due to~$D'$ will vary for different $D$. Robustness quantifies this difference in information gain across all possible data $D$.   We show that, amongst the beta distributions, the Jeffreys' prior exhibits the highest level of robustness.
	
The quantification of knowledge gained from additional data is a topic that has received limited attention in the literature. In the realm of foundational research on quantum theory, this issue has been acknowledged but not extensively explored. Summhammer initially proposed the notion that `more data from measurements lead to more knowledge about the system', but did not employ information theory to address this problem, instead using changes in measurement uncertainty to quantify knowledge obtained in the asymptotic limit.  This approach limits the applicability of the idea, as it excludes considerations pertaining to prior probability distributions and does not readily apply to finite data.
	
Wootters demonstrated the significance of Jeffreys' prior in the context of quantum systems from a different information-theoretical perspective~\cite{Wotters2013}. In the domain of communication through quantum systems, Jeffreys' prior can maximize the information gained from measurements. Wootters approaches the issue from a more systematic perspective, utilizing mutual information to measure the information obtained from measurements. However, mutual information quantifies the \emph{average information gain over all possible data sequences,} which is not suitable for addressing the specific scenario we discussed earlier, where the focus is on the information gain from a fixed data sequence.
	
More broadly, the question of how much information is gained with the acquisition of additional data has been a relatively under-explored topic in both practical applications and foundational research on quantum theory. Commonly, mutual information is employed as a utility function. However, as noted above, mutual information essentially represents the expected information gain averaged over all possible data sequences. Consequently, it does not address the specific question of how much information is gained when a particular additional data point is obtained. From our perspective, this averaging process obscures essential edge effects, including black swan events, which, as we will discuss, serve as valuable guides for selecting appropriate information measures.
	
While our investigation primarily focuses on information gain in quantum systems, the principles and conclusions we derive can be extended to general probabilistic systems characterized by fixed continuous parameters. Based on our analysis, we recommend quantification using differential information gain and the utilization of Jeffreys' prior. If one seeks to calculate the \textit{expected} information gain in the next step, both the expected differential information gain and the expected relative information gain can be employed since, as we demonstrate, they yield the same result.
	

	The paper is organized as follows.
	In Section \ref{sec:background}, we detail the two information gain measures, both of which have their origins in the generalization of Shannon entropy to continuous probability distributions. We will also delve into Jaynes' approach to continuous entropy, which serves as the foundation for understanding these two information gain measures.
	Sections \ref{sec:diff-finite-beta} and \ref{sec:rel-infor} focus on the numerical and asymptotic analysis of differential information gain and relative information gain. Our primary emphasis is on how these measures behave under different prior distributions. We will explore black swan events, where the additional data $D'$ is highly improbable given $D$. In this unique context, we will assess the physical meaningfulness of the two information gain measures.
	In Section \ref{sec:exp-diff}, we will discuss expected information gain under the assumption that data $D'$ from additional measurements has not yet been received. Despite the general differences between the two measures, it is intriguing to note that the two expected information gain measures are equal.
	Section \ref{sec:comparisons} presents a comparison of the two information gain measures and the expected information gain. It is within this section that we propose the \emph{Principle of Information Increase}, which crystallises the results of our analysis of the two information gain measures.
	Finally, Section \ref{sec:related-work} explores the relationships between our work and other research in the field.

	\section{Continuous Entropy and Bayesian Information Gain}
	\label{sec:background}
	
	\subsection{Entropy of Continuous Distribution}
	The Shannon entropy serves as a measure of uncertainty concerning a random variable before we have knowledge of its value. If we regard information as the absence of uncertainty, the Shannon entropy can also be used as a measure of information gained about a variable after acquiring knowledge about its value. However, it is important to note that Shannon entropy is applicable only to discrete random variables. To extend the concept of entropy to continuous variables, Shannon introduced the idea of differential entropy. Unlike Shannon entropy, differential entropy was not derived on an axiomatic basis.  Moreover, it has a number of  limitations.  
	
First, he differential entropy can yield negative values, as exemplified by the differential entropy of a uniform distribution over the interval $[0, \frac{1}{2}]$, which equals $-\!\log 2$. Negative entropy, indicating a negative degree of uncertainty, lacks meaningful interpretation.   Second, the differential entropy is coordinate-dependent~\cite{CoverThomas}, so that its value is not conserved under change of variables. This implies that viewing the same data through different coordinate systems may result in the assignment of different degrees of uncertainty.  Since the choice of coordinate systems is usually considered arbitrary, this coordinate-dependence also lacks a meaningful interpretation.
	
In an attempt to address the challenges associated with continuous entropy, Jaynes introduced a solution known as the limiting density of discrete points (LDDP) approach in his work \cite{Jaynes1963}. In this approach, the probability density $p(x)$ of a random variable $X$ is initially defined on a set of discrete points $x \in {x_1, x_2, \cdots, x_n}$. Jaynes proposed an invariant measure $m(x)$ such that, as the collection of points ${x_i}$ becomes increasingly numerous, in the limit as $n \rightarrow \infty$, 
	\begin{equation}
		\lim_{n\rightarrow \infty} \frac{1}{n}\text{~(number of points in~$a<x<b$)} = \int_a^b m(x) dx
	\end{equation}
	
	With the help of~$m(x)$, the entropy of~$X$ can then be represented as
	\begin{equation}
		\label{continuous_entropy}
		H(X) = \lim_{n\rightarrow \infty} \log{n} - \int p(x)\log\frac{p(x)}{m(x)}dx
	\end{equation}
	
	In this manner, the weaknesses associated with differential entropy appear to be resolved. This quantity remains invariant under changes of variables and is always non-negative. A similar approach is also discussed in~\cite{CoverThomas}. However, two new issues arise. In Eq.~(\ref{continuous_entropy}), $H(X)$ contains an infinite term, and the measure function $m(x)$ is unknown.
	
Regarding the infinite term, two potential solutions exist. The first option is to retain this infinite term, and to reserve interpretation to the \emph{difference} in the continuous entropy of two continuous distributions. The second solution is more straightforward, simply to omit the problematic~$\log n$ term.
	
	\begin{enumerate}
		\item Entropy of continuous distribution as a difference
		
		For example, when variable~$X$ is updated to~$X'$ due to certain actions, the \textit{decrease} in entropy can be expressed as:
		\begin{equation}
			\label{difference_entropy}
			X \rightarrow X', \quad \Delta H(X\rightarrow X') = H(X') - H(X) = \int p'(x)\log\frac{p'(x)}{m(x)}dx - \int p(x)\log\frac{p(x)}{m(x)}dx
		\end{equation}
		where $p'(x)$ represents the probability distribution of $X'$. In this context, we assume that the two infinite terms cancel. The quantity $\Delta H$ quantifies the reduction in uncertainty about variable $X$ resulting from these actions. This reduction in uncertainty can also be interpreted as an increase in information.
		
		\item Straightforward solution
		
		Jaynes directly discards the infinity term in Eq.~(\ref{continuous_entropy}). For the sake of convenience, the minus sign is also dropped. This leads to the definition of Shannon-Jaynes information:
		\begin{equation}
			\label{Jaynes_information}
			H_{Jaynes}(X) = \int p(x)\log\frac{p(x)}{m(x)}dx
		\end{equation}
		This term quantifies the amount of information we possess regarding the outcome of $X$ rather than the degree of uncertainty about $X$. $H_{Jaynes}$ is equivalent to the KL divergence between the distributions $p(x)$ and $m(x)$.
		
	\end{enumerate}
	
	In short, there are two ways to represent the entropy of continuous distribution, and there is no obvious criterion to choose between them. In a special case where the variable $X$ initially follows a distribution identical to the measure function, i.e., $p(x) = m(x)$, and $X$ undergoes evolution to $X'$ with distribution $p'(x)$, then we find that $\Delta H(X \rightarrow X') = H_{Jaynes}(X')$.
	
	The remaining challenge lies in the selection of the measure function $m(x)$. When applying this concept of continuous entropy to the relationship between information theory and statistical physics, Jaynes opted for a uniform measure function~\cite{Jaynes1963}. However, it is far from clear that this choice is universally applicable.  Currently, there is no established criterion for the choice of the measure function.  It is worth noting that this measure function acts analogously to the prior distribution in the context of Bayesian probability.
	
	\subsection{Bayesian Information Gain}
	
	In a coin-tossing model, let $p$ denote the probability of getting a head in a single toss, and $N$ be the total number of tosses. After $N$ tosses, the outcomes of these $N$ tosses can be represented by an $N$-tuple, denoted as $T_N = (t_1, t_2, \cdots, t_N)$, where each $t_i$ represents the result of the $i$th toss, with $t_i$ taking values in the set $\{\text{Head}, \text{Tail}\}$.
	Applying Bayes' rule, the posterior probability for the probability of getting a head is given by:
	\begin{equation}
		\Pr(p|N,T_N,I) =\frac{\Pr(T_N|N,p,I)\Pr(p|I)}{\int \Pr(T_N|N,p,I)\Pr(p|I) dp}
	\end{equation}
	where $\Pr(p|I)$ represents the prior. The information gain after~$N$ tosses would be the KL divergence from the prior distribution to the posterior distribution:
	\begin{equation}
		I(N) = D_\text{KL}( \Pr(p|N,T_N,I) || \Pr(p|I) )=\int^1_0 \Pr(p|N,T_N,I) \ln{\frac{\Pr(p|N,T_N,I)}{\Pr(p|I)}} dp
	\end{equation}

	Based on the earlier discussion on continuous entropy, this quantity can be interpreted in two ways, either as the difference between the information gain after~$N$ tosses and the information gain without any tosses; or as the KL divergence from the posterior distribution to the prior distribution.
	
	When considering the information gain of additional tosses based on the results of the previous $N$ tosses, we may observe two different approaches to represent this quantity.
	
	Let $t_{N+1}$ represent the outcome of the $(N+1)$th toss, and let $T_{N+1} = (t_1, t_2, \ldots, t_N, t_{N+1})$ denote the combined outcomes of the first $N$ tosses and the $(N+1)$th toss. The posterior distribution after these $N+1$ tosses is given by:
	\begin{equation}
		\Pr(p|N+1,T_{N+1},I) =\frac{\Pr(T_{N+1}|{N+1},p,I)\Pr(p|I)}{\int \Pr(T_{N+1}|{N+1},p,I)\Pr(p|I) dp}
	\end{equation}

	When considering information gain as a difference between two quantities, the first form of information gain for this single toss $t_{N+1}$ can be expressed as:
	\begin{equation}
		I_{\text{diff}} = D_\text{KL}(\Pr(p|N+1,T_{N+1},I)||\Pr(p|I)) - D_\text{KL}(\Pr(p|N,T_N,I)||\Pr(p|I))
	\end{equation}

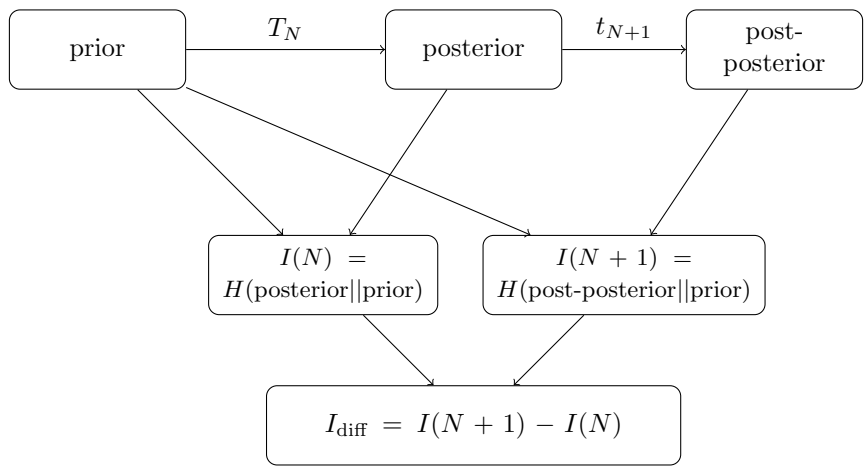
\begin{figure}[!h]
		\begin{center}
			\begin{tikzpicture}
				\node[block] (prior) at(0,0) {prior};
				\node[block] (posterior) at(5 ,0) {posterior};
				\node[block] (post-posterior) at(9,0) {post-posterior};
				\node[medium_block](deltapost) at(3,-3) {\small $I(N) = H(\text{posterior}||\text{prior})$};
				\node[mb_block](deltapostpost) at(7,-3) {\small $I(N + 1) = H(\text{post-posterior}||\text{prior})$};
				\node[big_block](delta) at(5,-5) {$ I_{\text{diff}} = I(N+1) - I(N)$};
				\draw[->] (prior) --node[above]{$T_N$}(posterior);
				\draw[->] (posterior) --node[above]{$t_{N+1}$}(post-posterior);
				\draw[->] (prior) --(deltapost);
				\draw[->] (posterior) -- (deltapost);
				\draw[->] (prior) --(deltapostpost);
				\draw[->] (post-posterior) --(deltapostpost);
				\draw[->] (deltapost) --(delta);
				\draw[->] (deltapostpost) --(delta);
			\end{tikzpicture}
		\end{center}
		\caption{\label{fig:differential-information-gain}\textit{Differential Information Gain in a Single Toss.} Assuming we have data from the first $N$ tosses, denoted as $T_N$. Using a specific prior distribution, we can calculate the information gain for these first $N$ tosses, denoted as $I(N)$. If we now consider the $(N+1)$th toss and obtain the result $t_{N+1}$, we can repeat the same procedure to calculate the information gain for a total of $N+1$ tosses, denoted as $I(N+1)$. The information gain specific to the $(N+1)$th toss can be obtained as the difference between $I(N+1)$ and $I(N)$.}
	\end{figure}

	In this expression, the first term $H(\Pr(p|N+1, t_{N+1}, I) || \Pr(p|I))$ represents the information gain from $0$ tosses to $N+1$ tosses, while the second term $H(\Pr(p|N, T_N, I) || \Pr(p|I))$ represents the information gain from $0$ tosses to $N$ tosses. The difference between these terms quantifies the information gain in the single $(N+1)$th toss~(see Fig.~\ref{fig:differential-information-gain}). In this context, we can refer to $I_{\text{diff}}$ as the \emph{differential information gain in a single toss.}
		
\begin{figure}[!h]
		\begin{center}
			\begin{tikzpicture}
				\node[block] (prior) at(0,0) {prior};
				\node[block] (posterior) at(5 ,0) {posterior};
				\node[block] (post-posterior) at(9,0) {post-posterior};
				
				\node[big_block](deltapostpost) at(7,-3) {\small $I_{\text{rel}} = H(\text{post-posterior}||\text{posterior})$};

				\draw[->] (posterior) --node[above]{$t_{N+1}$}(post-posterior);
				\draw[->] (prior) --node[above]{$T_N$}(posterior);
				\draw[->] (posterior) -- (deltapostpost);
				
				\draw[->] (post-posterior) --(deltapostpost);

			\end{tikzpicture}
		\end{center}
		\caption{\label{fig:relative-information-gain}\textit{Relative Information Gain in a Single Toss.} The posterior distribution calculated from the results of the first $N$ tosses serves as the prior for the $(N+1)$th toss. The KL divergence between this posterior and the subsequent posterior represents the information gain in the $(N+1)$th toss.}
	\end{figure}
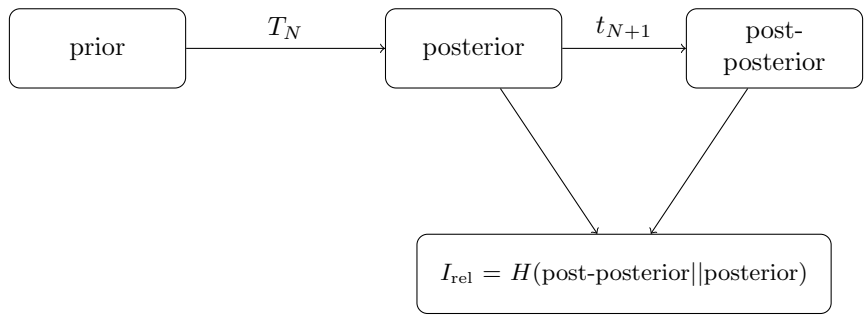

	Alternatively, we can adopt a straightforward approach of directly calculate the information gain specifically from the $N$th toss to the $(N+1)$th toss.  Hence, the second form of information gain is defined as follows:
	\begin{equation}
		I_{\text{rel}} = D_\text{KL}(\Pr(p|N+1,T_{N+1},I)||\Pr(p|N,T_N,I)),
	\end{equation}
which is simply the KL divergence from the posterior distribution after $N$ tosses to the posterior distribution after $N+1$ tosses~(see Fig.~\ref{fig:relative-information-gain}).
  	In this case, we refer to $I_{\text{rel}}$ as the \emph{relative information gain in a single toss.}
	
	In general, these two quantities, $I_{\text{diff}}$ and $I_{\text{rel}}$, are not the same, unless $N=0$, which implies that no measurements have been performed. $I_{\text{diff}}$ could take on negative values, while $I_{\text{rel}}$ is always non-negative due to the properties of the KL divergence. Although KL divergence is not a proper distance metric between probability distributions (as it does not satisfy the triangle inequality), it is a valuable tool for illustrating the analogy of displacement and distance in a random walk model. This analogy helps elucidate the subtle difference between the two types of information gain.
	
	We aim to determine which information gain measure is a more suitable choice and introduce an informational postulate to guide our decision. This postulate comes from an intuitive idea: `more measurements lead to more knowledge about the physical system' \cite{Summhammer1994, Summhammer1999}. For instance, when measuring the value of a physical quantity, we often perform multiple identical measurements to reduce statistical fluctuations in the values. We contemplate whether this idea can be reformulated in terms of information theory. If we quantify `knowledge' in terms of information gain from data, this notion suggests that the information gain from additional data should be positive if it indeed contributes to our understanding.
	
	This consideration makes relative information gain an appealing choice, as it is always non-negative. However, the derivation of differential information gain also carries significance. This leads to the question of whether this intuitive idea has physical meaning, and if not, what might be a reasonable interpretation of it.
	In the following sections, we will delve into the concept of differential information gain, both in finite $N$ cases and asymptotic cases. We will explore the implications of negative values of information gain, particularly in extreme situations. Additionally, we will conduct numerical and asymptotic analyses of relative information gain.
	After analyzing both information gain measures, we will be better equipped to compare and establish connections between them, and to assess the physical meaningfulness of the intuitive idea we set out to explore.

	\section{Differential Information Gain}
	\label{sec:diff-finite-beta}
	
	\subsection{Finite number of tosses}
	The prior distribution we employ is the beta distribution, which serves as the conjugate prior for the binomial distribution.
	
	\begin{equation}
		\label{prior}
		\Pr(p|I)=\frac{p^{\alpha}(1-p)^{\alpha}}{B(\alpha+1,\alpha+1)}
	\end{equation}
	where~$\alpha>-1$, and $B(\cdot,\cdot)$ is the beta function.
	
	In general, the beta distribution is characterized by two parameters, and for the sake of convenience, we work with a simplified single-parameter beta distribution. This single-parameter beta distribution encompasses a wide spectrum of priors, including the uniform distribution (when $\alpha = 0$) and Jeffreys' prior (when $\alpha = -0.5$).
	
	The differential information gain of the $(N+1)$th toss is (see Appendix~\ref{appendix:derivation_dig})
	\begin{equation}
		\label{dig_formula}
		\begin{aligned}
			I_{\text{diff}}=&~\psi(h_N+\alpha+2)-\psi(N+2\alpha+3)\\ 
			&~+\frac{h_N}{h_N+\alpha+1}-\frac{N}{N+2\alpha+2}
			+\ln{\frac{N+2\alpha+2}{h_N+\alpha+1}}
		\end{aligned}
	\end{equation}
	In this context, we assume that $t_{N+1} = \text{`Head'}$. It's worth noting that there is also a corresponding $I_{\text{diff}}(t_{N+1} = \text{`Tail'})$, but for brevity, we will focus on a single case. Our calculations consider all possible values of $T_N$, and the expressions for both cases (Head and Tail) are symmetric.
	
	$I_{\text{diff}}$ is a function of $h_N$ and $\alpha$, and $h_N$ ranges from $0$ to $N$. We will select a specific value for $\alpha$ and calculate all the~$N+1$ values of~$I_{\text{diff}}$ for each~$N$.
	
	\subsubsection{Positivity of $I_{\text{diff}}$} 
	Returning to our initial question: ``Will more data leads to more knowledge?".  If we use the term ``knowledge" to represent the differential information gain and use $I_{\text{diff}}$ to quantify the information gained in each measurement, the question becomes rather straightforward: ``Is $I_{\text{diff}}$ always positive?"
	
\begin{figure}[!h]		
			\includegraphics{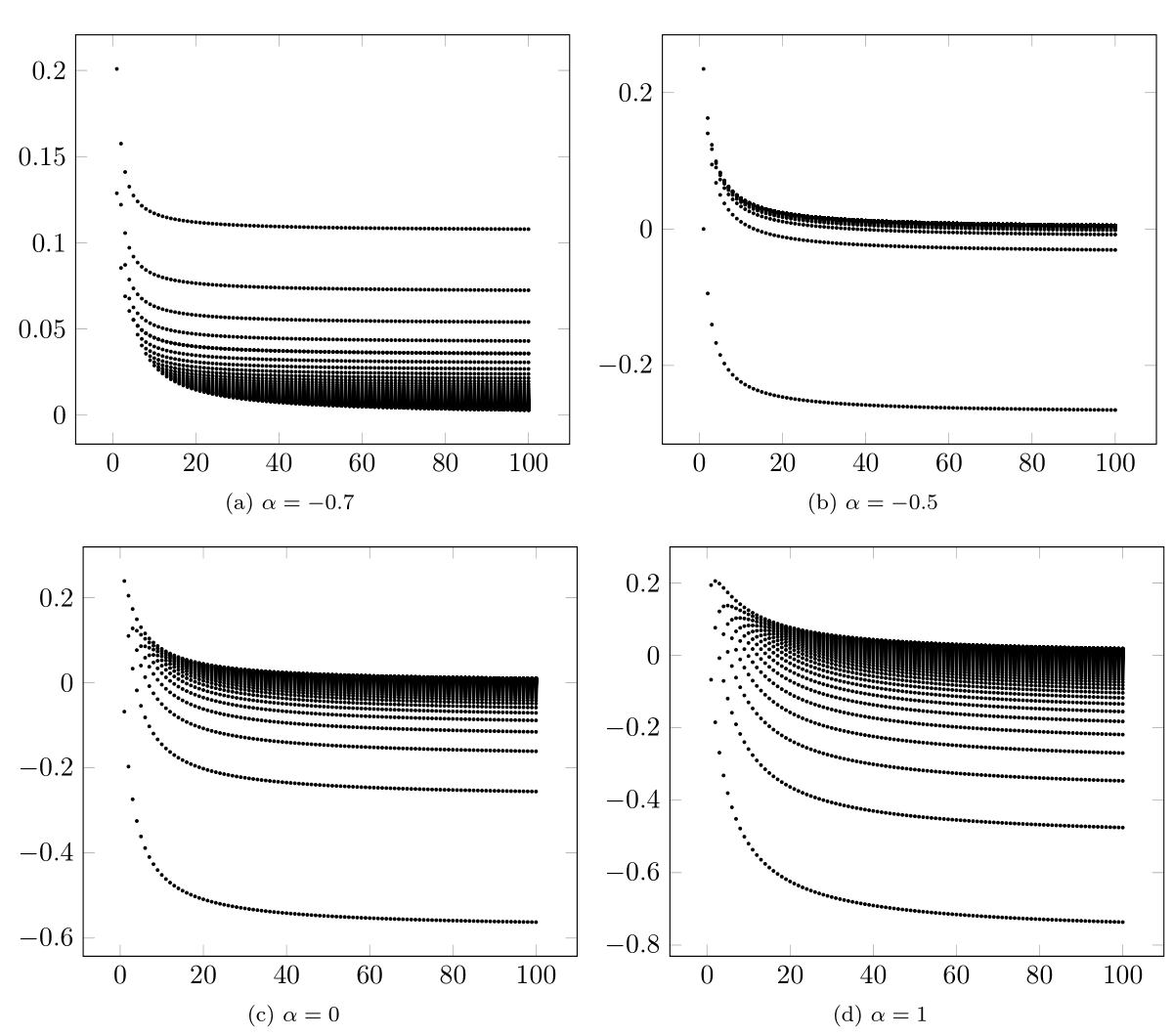}
		\caption{\label{dig_graph} \textit{Differential Information Gain ($I_{\text{diff}}$) vs. $N$ for Different Priors.} Here, the $y$-axis represents the value of $I_{\text{diff}}$, and the $x$-axis corresponds to the value of $N$. In each graph, we fix the value of $\alpha$ to allow for a comparison of the behavior of $I_{\text{diff}}$ under different priors. Given $N$, there are $N+1$ points as $h_N$ ranges from $0$ to $N$. Notably, for $\alpha = -0.7$, all points lie above the $x$-axis, while for other priors, negative points are present and the fraction of negative points becomes constant as $N$ increases. The asymptotic behavior of this fraction will be shown in Figure \ref{FoNvsN_graph}. Moreover, it appears that the graph is most concentrated when $\alpha = -0.5$, whereas for $\alpha < -0.5$ and $\alpha > -0.5$, the graph becomes more dispersed. This dispersive/concentrating feature is clearly depicted in Figure \ref{dig_stddev}.}
	\end{figure}

	In Figure \ref{dig_graph}, we present the results of numerical calculations for various values of $N$. Upon close examination of the graph, it becomes evident that the $I_{\text{diff}}$ is not always positive, except under specific conditions. In the following sections, we will delve into the conditions that lead to exceptions.
	
	For certain priors, the differential information gain is consistently positive (Figure \ref{dig_graph}a); while for other priors, both positive and negative regions exist (Figure \ref{dig_graph}b, \ref{dig_graph}c, \ref{dig_graph}d). It's worth noting that for priors leading to negative regions, the lowest line exhibits greater dispersion compared to the other data lines. This lower line represents the scenario where the first $N$ tosses all result in tails, but the $(N+1)$th toss yields a head. This situation is akin to a black swan event, and negative information gain in this extreme case holds significant meaning---if we have tossed a coin $N$ times and obtaining all tails, we anticipate another tail in the next toss; hence, receipt of a heads on the next toss raises the degree of uncertainty about the outcome of the next toss, leading to a reduction in information about the coin's bias.
	
	\subsubsection{Fraction of Negatives}
	
	In order to illustrate the variations in the positivity of information gain under different priors, we introduce a new quantity known as the Fraction of Negatives (FoN), which represents the ratio of the number of $h_N$ values that lead to negative $I_{\text{diff}}$ and $N+1$. For instance, if, for a given $\alpha$, $N=10$ and~$I_{\text{diff}}<0$ when~$h_N=0,1,2,3$, the FoN under this~$\alpha$ and~$N$ is $\frac{4}{11}$.
	
	\begin{figure}[!h]
		\centering
			\includegraphics{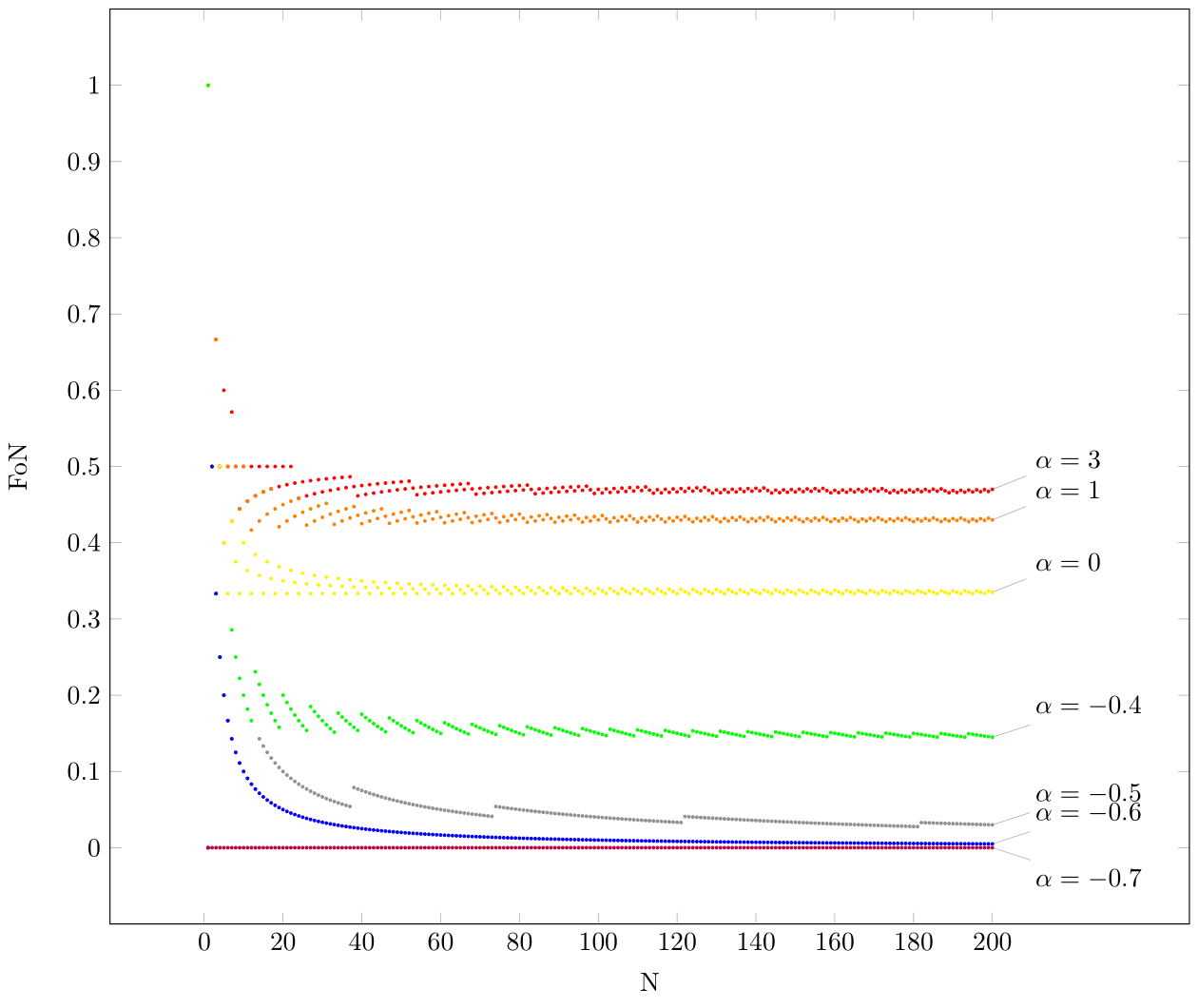}	
		\caption{\label{FoNvsN_graph} \textit{Fraction of Negatives~(FoN) vs. $N$ under different $\alpha$.} In Figure \ref{dig_graph}, we can observe that larger~$\alpha$ values lead to more dispersed lines and an increased number of negative values for each~$N$. We use FoN to quantify this fraction of negative points. It appears that for $\alpha \le -0.7$, FoN is consistently zero, indicating that $I_{\text{diff}}$ is always positive. For $\alpha \le -0.5$ FoN decreases and tends to be zero as $N$ becomes large, while for $\alpha > -0.5$, FoN tends to a constant as $N$ increases, and this constant grows with increasing $\alpha$.}
	\end{figure}
	
	From Figure \ref{FoNvsN_graph}, we identify a critical point, denoted as $\alpha_p$, which is approximately $-0.7$. For any $\alpha \le \alpha_p$, $I_{\text{diff}}$ is guaranteed to be positive for all $N$ and $h_N$ values.
	
	If $\alpha > \alpha_p$, negative terms exist for some $h_N$; however, the patterns of these negative terms differ across various $\alpha$ values.
	
	Additionally, we notice the presence of a turning point, $\alpha_0 = -0.5$. For $\alpha \le \alpha_0$, FoN tends to zero as $N$ increases, whereas for $\alpha > \alpha_0$, FoN approaches a constant as $N$ grows.
	
\begin{figure}[!h]
				\includegraphics[width=16cm]{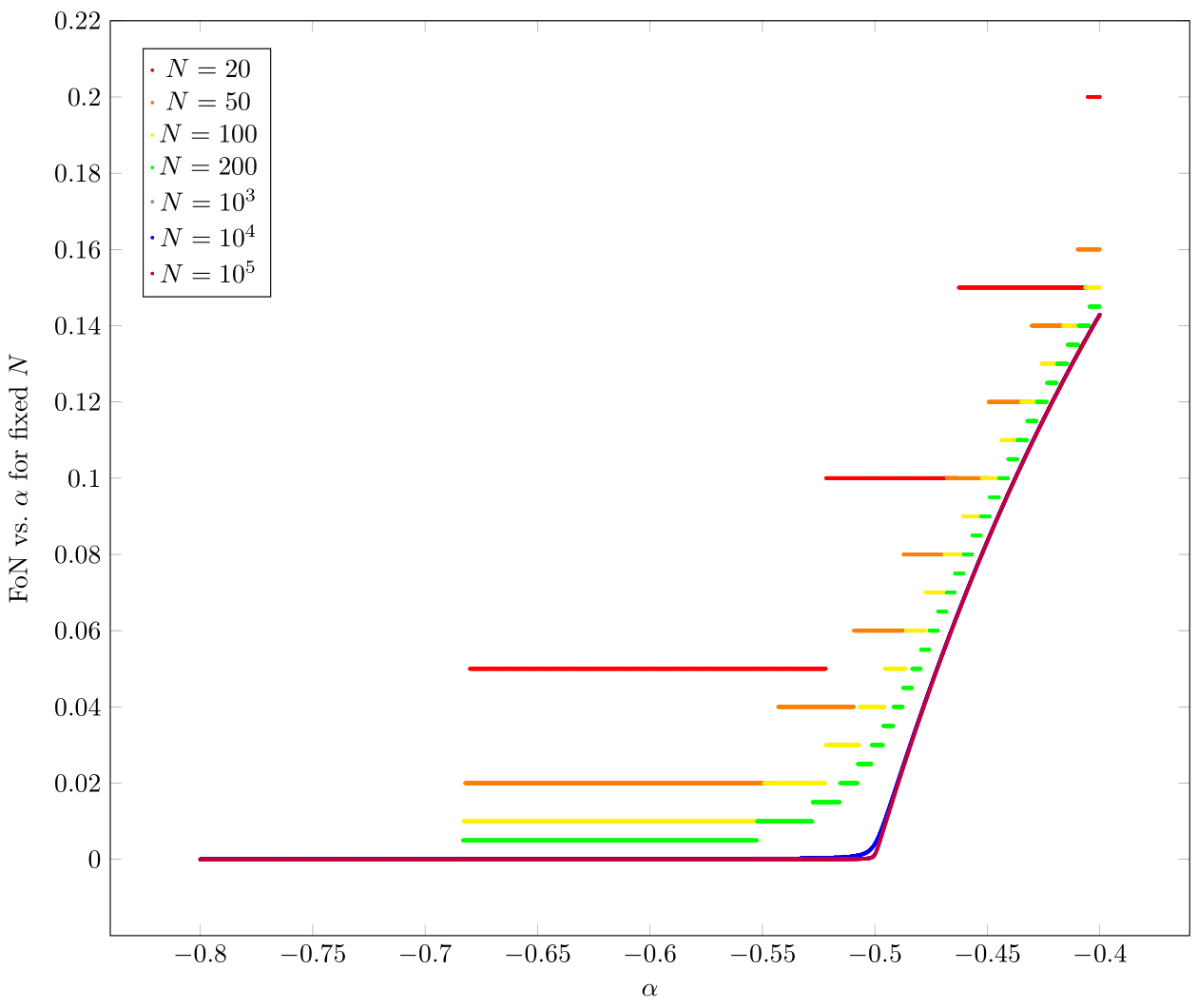}
		\caption{\label{FoNvsAlpha_graph} \textit{Fraction of Negatives (FoN) vs. $\alpha$ for Different~$N$.} We identify a critical point, denoted as $\alpha_p$, where the FoN equals zero when $\alpha \le \alpha_p$. The critical point exhibits a gradual variation with respect to $N$, following these patterns: (i) for small $N$, $\alpha_p$ is close proximity to~$-0.68$; (ii) for large $N$, $\alpha_p$ tends to~$-0.5$.}
	\end{figure}

	A clearer representation of the critical point $\alpha_p$ and the turning point $\alpha_0$ can be found in Figure \ref{FoNvsAlpha_graph}, where the critical point~$\alpha_p$ is approximately $-0.68$.

	\subsubsection{Robustness of $I_{\text{diff}}$}
\begin{figure}[!h]
				\includegraphics[width=16cm]{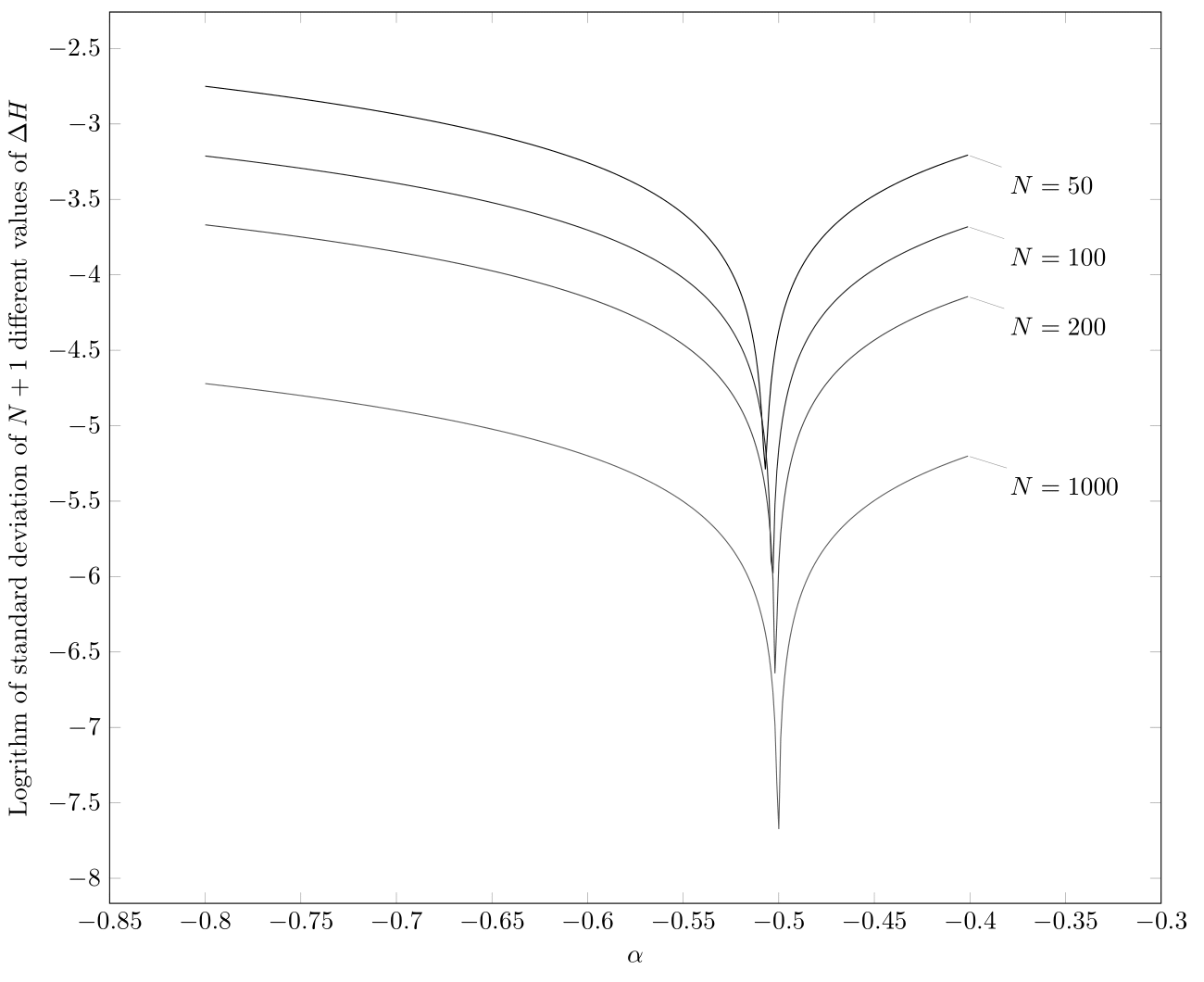}
		\caption{\label{dig_stddev} \textit{Robustness of Differential Information Gain ($I_{\text{diff}}$).} The $y$-axis represents the logarithm of the standard deviation of~$I_{\text{diff}}$ over all possible $h_N$ values, while the $x$-axis depicts various selections of $\alpha$. A smaller standard deviation indicates that different $h_N$ values lead to the same result, implying greater independence of $I_{\text{diff}}$ from $h_N$. This independence signifies the robustness of $I_{\text{diff}}$ with respect to the natural variability in $h_N$, as we consider $h_N$ to be solely determined by nature. The standard deviation, given a fixed $N$, is notably influenced by $\alpha$, and there exists an $\alpha$ value at which the dependence on $h_N$ is minimized. This particular $\alpha$ value approaches $-0.5$ as $N$ increases.}
	\end{figure}

	In Figure \ref{dig_graph}, different priors not only exhibit varying degrees of positivity but also display varying degrees of variation in $I_{\text{diff}}$ for different values of $h_N$, to which we refer as \emph{divergence}.  The divergence depends upon the choice of prior.  To better understand this dependence, we quantify the dependence of $I_{\text{diff}}$ on $h_N$ by the standard deviation of $I_{\text{diff}}$ across different values of $h_N$. Figure \ref{dig_stddev} illustrates how the standard deviation changes with respect to $\alpha$ while keeping $N$ constant.
	
	It is evident that when $\alpha$ is close to $-0.5$, the standard deviation is at its minimum. Reduced dependence of $I_{\text{diff}}$ on $h_N$ enhances its robustness against the effects of nature, as we attribute $h_N$ to natural factors while $N$ is determined by human measurement choices. As $N$ increases, the minimum point approaches $-0.5$. In the limit of large $N$, this minimum point will eventually converge to $\alpha=-\frac{1}{2}$, which means that, under this specific choice of prior, $I_{\text{diff}}$ depends minimally on $h_N$ and primarily on $N$.
	
	\subsection{Large~$N$ approximation}
	\label{subsec:diff-asymp}
	
	Utilizing a recurrence relation and the large $x$ approximation, the digamma function can be approximated as follows:
	\begin{equation}
		\psi(x) = \frac{1}{x-1} + \psi(x-1) \approx \frac{1}{x-1} + \ln(x-1) - \frac{1}{2(x-1)} = \frac{1}{2(x-1)}+ \ln(x-1)
	\end{equation}
	
	As a result, the large $N$ approximation for the differential information gain in Eq.~(\ref{dig_formula}) becomes:
	\begin{equation}
		\label{large_N_delta_H}
		I_{\text{diff}} = \frac{2h_N+1}{2(h_N+\alpha+1)} - \frac{2N+1}{2(N+2\alpha+2)}
	\end{equation}
	
	It is evident that when $\alpha=-\frac{1}{2}$, $I_{\text{diff}} =\frac{1}{2(N+1)}$, indicating that $I_{\text{diff}}$ solely depends on $N$. This finding aligns with Figure \ref{dig_graph}, which demonstrates that $I_{\text{diff}}$ is most concentrated when $\alpha = -0.5$, and is consistent with the results of \cite{Goyal2005}.
	
	In Figure \ref{FoNvsN_graph}, we observe that the FoN tends to become constant for very large $N$. These constants can be estimated using the large $N$ approximations of $I_{\text{diff}}$ in Eq.~(\ref{large_N_delta_H}). If $I_{\text{diff}} \le 0$, then
	\begin{equation}
		h_N \le \frac{2N\alpha+N+\alpha+1}{4\alpha+3},
	\end{equation}
and 	we  obtain:
	\begin{equation}
		\text{FoN} = \frac{1}{N+1}\frac{2N\alpha+N+\alpha+1}{4\alpha+3} \approx \frac{2\alpha+1}{4\alpha+3}
	\end{equation}
	This equation aligns with the asymptotic lines in Figure \ref{FoNvsN_graph}, providing support for the observation mentioned in Figure \ref{dig_graph}, namely that, for $\alpha = -0.7$, all points lie above the $x$-axis, while for other priors, negative points are present and the fraction of negative points becomes constant.
	
	\begin{table}[!h]
		\begin{center}{\scriptsize
			\begin{tabular}{|p{1cm}|p{4.2cm}|p{4cm}|p{4cm}|}
				\hline
				$\alpha$ &  FoN~(numerical result,~$N=1000$) &	FoN~(asymptotic result) & discrepancy between two results\\ \hline
				-0.7	&	0	&	0	&	0\\ \hline
				-0.6	&0.001	&	0	&$0.1\%$\\ \hline
				-0.5	&0.013	&	0	&$1.3\% $\\ \hline
				-0.4	&0.144	&0.143	&$ 0.1\%$\\ \hline
				0		&0.334	&0.333	&$ 0.1\%$\\	\hline
				1		&0.429	&0.429	&0\\	\hline
				3		&0.467	&0.467	&0\\
				\hline
			\end{tabular}}
		\end{center}
		\caption{\textit{Fraction of Negatives (FoN) under Selected Priors.} A comparison between numerical results and asymptotic results demonstrates their agreement.}
	\end{table}

	\section{Relative Information Gain}
	\label{sec:rel-infor}
	
	The second form of information gain in a single toss is relative information gain, which represents the KL divergence from the posterior after $N$ tosses to the posterior after $N+1$ tosses.	We continue to use the one-parameter beta distribution prior in the form of Eq.~(\ref{prior}). The relative information gain is~(see appendix~\ref{appendix:derivation_rig}):
	\begin{equation}
		\label{rig_formula}
		I_{\text{rel}}(t_{N+1}={\text{`Head'}}) = \psi(h_N+\alpha+2)-\psi(N+2\alpha+3)+\ln{\frac{N+2\alpha+2}{h_N+\alpha+1}}
	\end{equation}
	
	Relative information gain exhibits entirely different behavior compared to differential information gain. Due to the properties of KL divergence, relative information gain is always non-negative, eliminating the need to consider negative values. We aim to explore the dependence of relative information gain on priors and the interpretation of information gain in extreme cases.
	
	\begin{figure}[!h]
			\includegraphics[width=16cm]{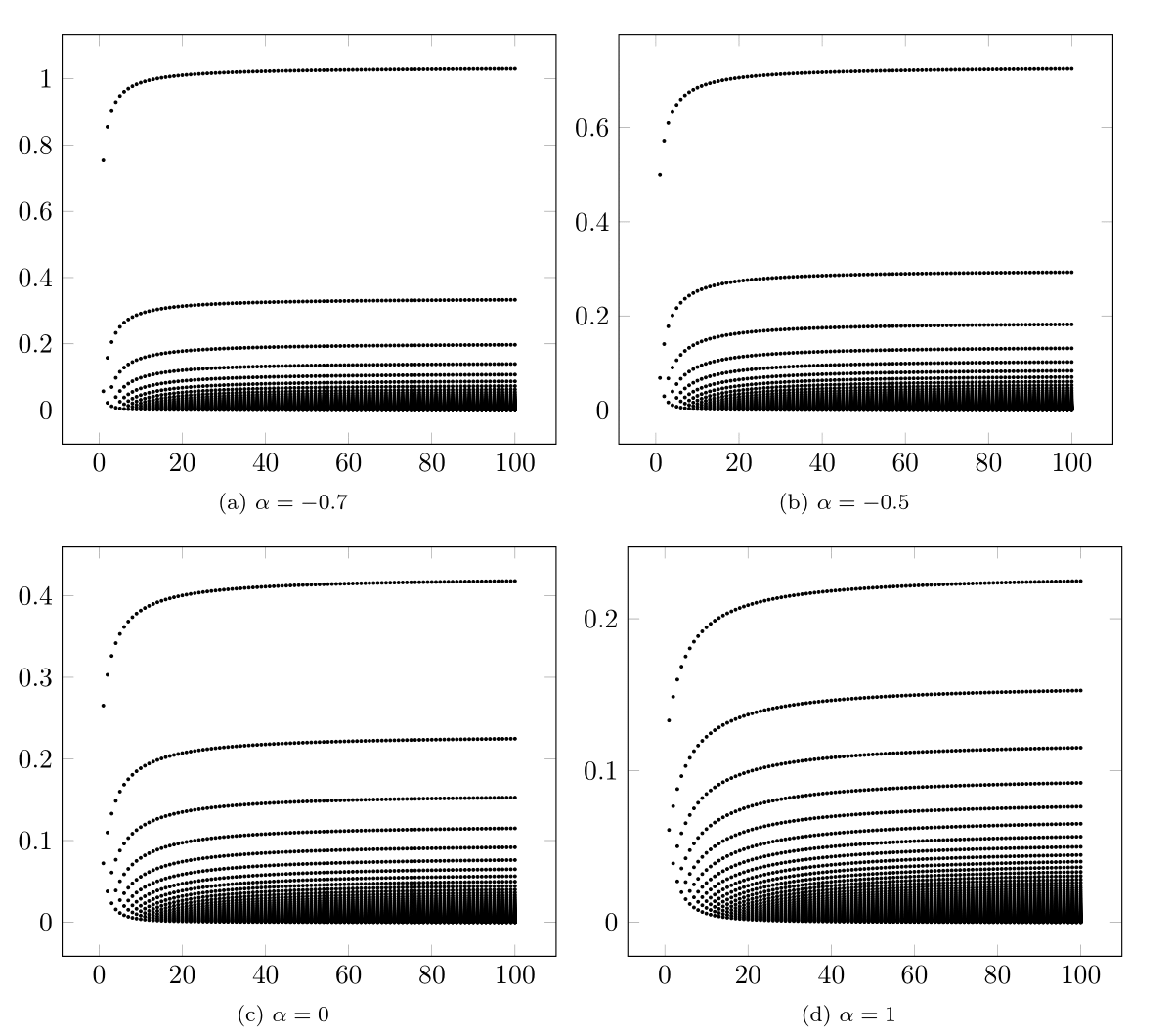}
		\caption{\label{rig_graph} \textit{Relative Information Gain ($I_{\text{rel}}$) over Different Priors.} The $y$-axis represents the value of $I_{\text{rel}}$, while the $x$-axis represents $N$. For each~$N$ there are~$N+1$ different values of~$I_{\text{rel}}$. It is important to note that $I_{\text{rel}}$ is consistently positive across these selected priors. Similar to the differential information gain, each graph displays numerous divergent lines. However, the shape of these divergent lines remains remarkably consistent across varying values of $\alpha$. The majority of these lines fall within the range of $I_{\text{rel}}$ between $0$ and $0.2$.}
	\end{figure}
	
	In Figure \ref{rig_graph}, it becomes evident that, under different priors, the data lines exhibit similar shapes. This suggests that relative information gain is relatively insensitive to the choice of priors. On each graph, the top line represents the extreme case where the first $N$ tosses result in tails, and the $(N+1)$th toss results in a head. This line is notably separated from the other data lines, indicating that relative information gain behaves more like a measure of the degree of surprise associated with this additional data. In this black swan event,  the posterior after $N+1$ tosses differs significantly from the posterior after $N$ tosses.
	
	\begin{figure}[!h]
			\includegraphics[width=16cm]{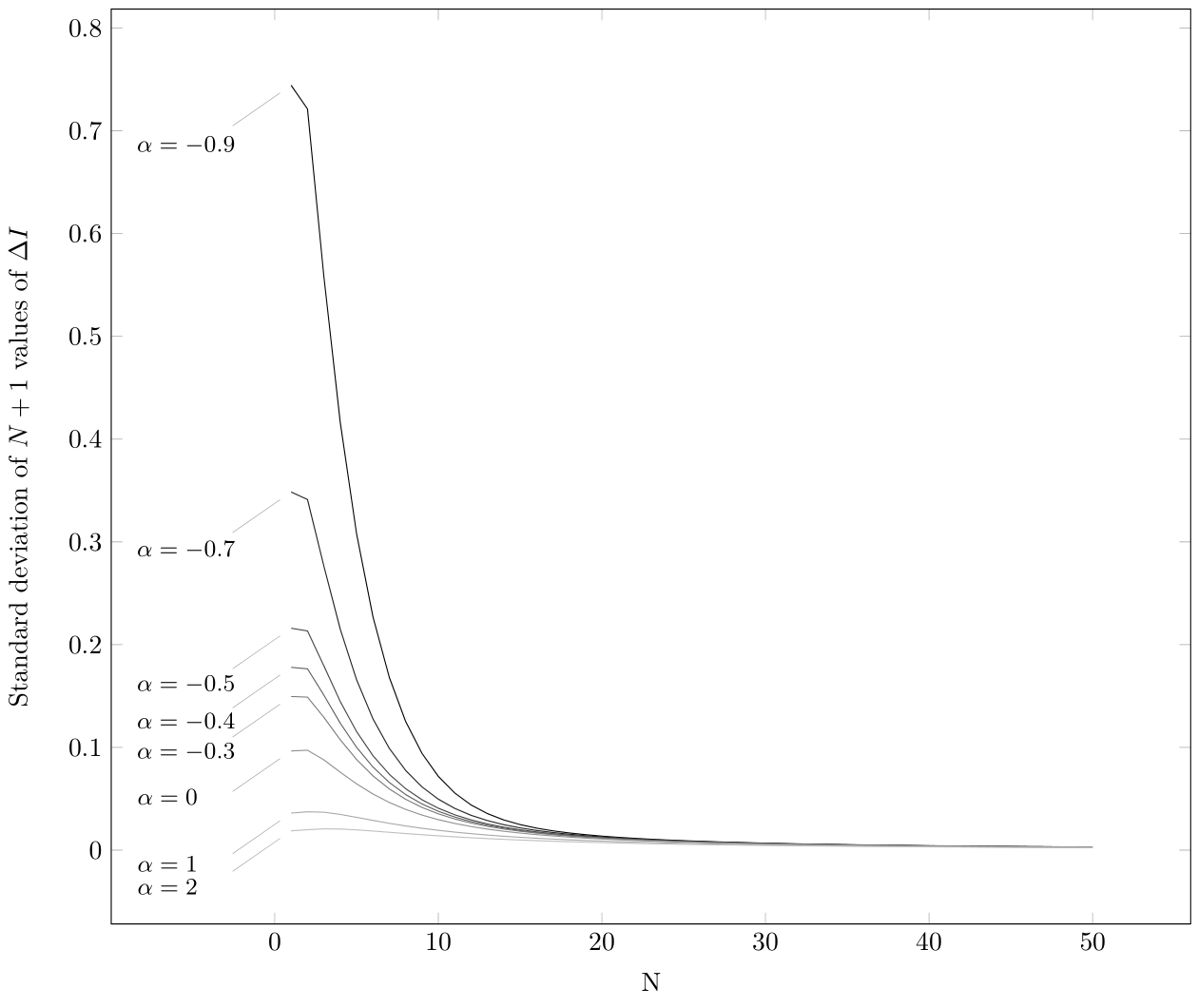}		
		\caption{\label{rig_stddev} \textit{Robustness of Relative Information Gain ($I_{\text{rel}}$)}. The $y$-axis represents the standard deviation of~$I_{\text{rel}}$ across all possible~$h_N$. This demonstrates the substantial independence of $I_{\text{rel}}$ from $h_N$. Additionally, as $N$ increases, the standard deviations tend to approach zero for all priors.}
	\end{figure}
	
	For small $N$, both the average value and standard deviation of $I_{\text{rel}}$ exhibit a clear monotonic relationship with $\alpha$, meaning that larger values of $\alpha$ result in smaller average values and standard deviations. However, as $N$ becomes large, all priors converge and become indistinguishable. Nonetheless, it is important to note that relative information gain remains heavily dependent on the specific data sequences ($h_N$).
	By utilizing the approximation of the digamma function, we can obtain:
	\begin{equation}
		\begin{aligned}
			I_{\text{rel}}(t_{N+1}=\text{`Head'}) &\approx \frac{1}{2(h_N+\alpha+1)}-\frac{1}{2(N+2\alpha+2)} = \frac{N-h_N+\alpha+1}{2(h_N+\alpha+1)(N+2\alpha+2)} \\
			&=\frac{\frac{\alpha}{N}+1-\frac{h_N}{N}}{2(\frac{h_N}{N}+\frac{\alpha+1}{N})(N+2\alpha+2)}			
		\end{aligned}
	\end{equation}
	
	In the large $N$ limit, $I_{\text{rel}}$ becomes:
	\begin{equation}
		I_{\text{rel}}(t_{N+1}=\text{`Head'}) \approx \frac{N-h_N}{2h_N N}
	\end{equation}
	Thus, it appears that the properties of relative information gain and differential information gain are complementary to each other.  The differences between them are summarized in Table~\ref{tbl:information-gain-comparison}.
		\begin{table}[!h]
			\begin{center}{\scriptsize
				\begin{tabular}{|p{5cm}|p{5cm}|p{5cm}|}
					\hline
					&	Asymptotic forms ($t_{N+1}=\text{`Head'}$) &	Asymptotic sensitivity about prior \\ \hline
					Diff Infor. Gain & $I_{\text{diff}} \approx \frac{2h_N+1}{2(h_N+\alpha+1)} - \frac{2N+1}{2(N+2\alpha+2)}$ & Heavily dependent on prior and for some certain prior $I_{\text{diff}}$ will be independent of $h_N$ \\ \hline
					Rel Infor. Gain & $I_{\text{rel}} \approx \frac{N-h_N}{2h_N N}$ & Insensitive to prior and for large $N$ only affected by $h_N$	\\
					\hline
				\end{tabular}}
			\end{center}
	\caption{\label{tbl:information-gain-comparison} Comparison of characteristics of two measures of information gain.}
	\end{table}
	
	\section{Expected Information Gain}
	\label{sec:exp-diff}
	
	In this section we discuss a new scenario: After $N$ tosses, but before the $(N+1)$th toss has been taken, can we predict how much information gain will occur in the next toss? The answer is affirmative, as discussed earlier.
	
	After $N$ tosses, we obtain a data sequence $T_N$ with $h_N$ heads. However, we can only estimate the probability $p$ based on the posterior $\Pr(p|N,T_N,I)$. The expected value of $p$ can be expressed as:
	\begin{equation}
		\label{ave_p}
		\langle p \rangle = \int^1_0 p~\Pr(p|N,T_N,I)~dp = \frac{h_N+\alpha+1}{N+2\alpha+2}
	\end{equation}
	
	Based on this expected value of $p$, we can calculate the average of the information gain in the $(N+1)$th toss. We define the expected differential information gain in the $(N+1)$th toss as:
	\begin{equation}
		\label{exp_dig}
		\begin{aligned}
			\overline{I_{\text{diff}}}  =&~ \langle p \rangle \times I_{\text{diff}}(t_{N+1}=\text{`Head'}) + \langle 1-p \rangle \times I_{\text{diff}}(t_{N+1}=\text{`Tail'})\\
			=&~ \frac{h_N+\alpha+1}{N+2\alpha+2}\psi(h_N+\alpha+2)+\frac{N-h_N+\alpha+1}{N+2\alpha+2}\psi(N-h_N+\alpha+2)-\psi(N+2\alpha+3)\\
			&~+ \frac{h_N+\alpha+1}{N+2\alpha+2}\ln{\frac{N+2\alpha+2}{h_N+\alpha+1}} + \frac{N-h_N+\alpha+1}{N+2\alpha+2}\ln{\frac{N+2\alpha+2}{N-h_N+\alpha+1}}
		\end{aligned}
	\end{equation}
	
	$\overline{I_{\text{diff}}}$ represents the expected value of differential information gain in the $(N+1)$th toss. Similarly, we can define the expected relative information gain as:
	\begin{equation}
		\label{exp_rig}
		\begin{aligned}
			\overline{I_{\text{rel}}}  =&~ \langle p \rangle \times I_{\text{rel}}(t_{N+1}=\text{`Head'}) + \langle 1-p \rangle \times I_{\text{rel}}(t_{N+1}=\text{`Tail'}) \\
			=&~ \frac{h_N+\alpha+1}{N+2\alpha+2}\psi(h_N+\alpha+2)+\frac{N-h_N+\alpha+1}{N+2\alpha+2}\psi(N-h_N+\alpha+2)-\psi(N+2\alpha+3)\\
			&~+ \frac{h_N+\alpha+1}{N+2\alpha+2}\ln{\frac{N+2\alpha+2}{h_N+\alpha+1}} + \frac{N-h_N+\alpha+1}{N+2\alpha+2}\ln{\frac{N+2\alpha+2}{N-h_N+\alpha+1}}
		\end{aligned}
	\end{equation}
	
	Surprisingly, $\overline{I_{\text{diff}}}=\overline{I_{\text{rel}}}$. This relationship holds true for any prior, not being limited to the beta distribution type prior. Please refer to appendix~\ref{appendix:equivalence_exp} for a detailed proof. This suggests that there is only one choice for the expected information gain.
	
	\begin{figure}[!h]
			\includegraphics[width=16cm]{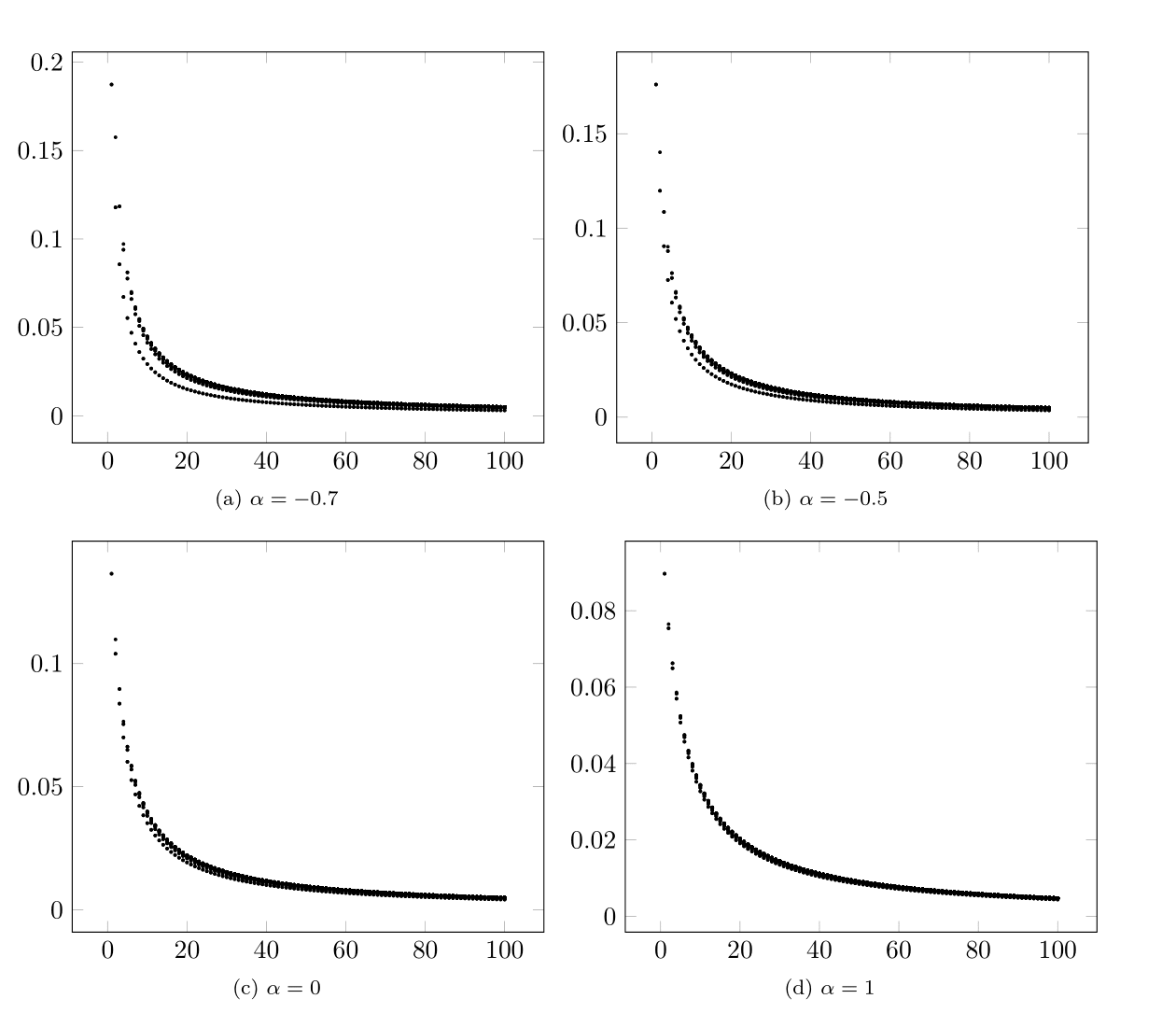}
		\caption{\label{expect_graph} \textit{Expected Information Gain vs.~$N$ for Fixed $\alpha$.} The $y$-axis represents the value of expected information, while the $x$-axis represents the value of~$N$. Notably, all expected information gain values are positive. The shapes of each graph exhibit remarkable similarity with a limited number of divergent lines. As $\alpha$ increases, the number of divergent lines decreases.}
	\end{figure}
	
	We first show the numerical results of expected information gain under different priors. It is evident that all data points are above the $x$-axis, indicating that the expected information gain is positive-definite, as anticipated. Since both $I_{\text{rel}}$ and $\langle p \rangle$ are positive, it follows that $\overline{I_{\text{rel}}}$ must also be positive.
	
\begin{figure}[!h]		
			\includegraphics[width=16cm]{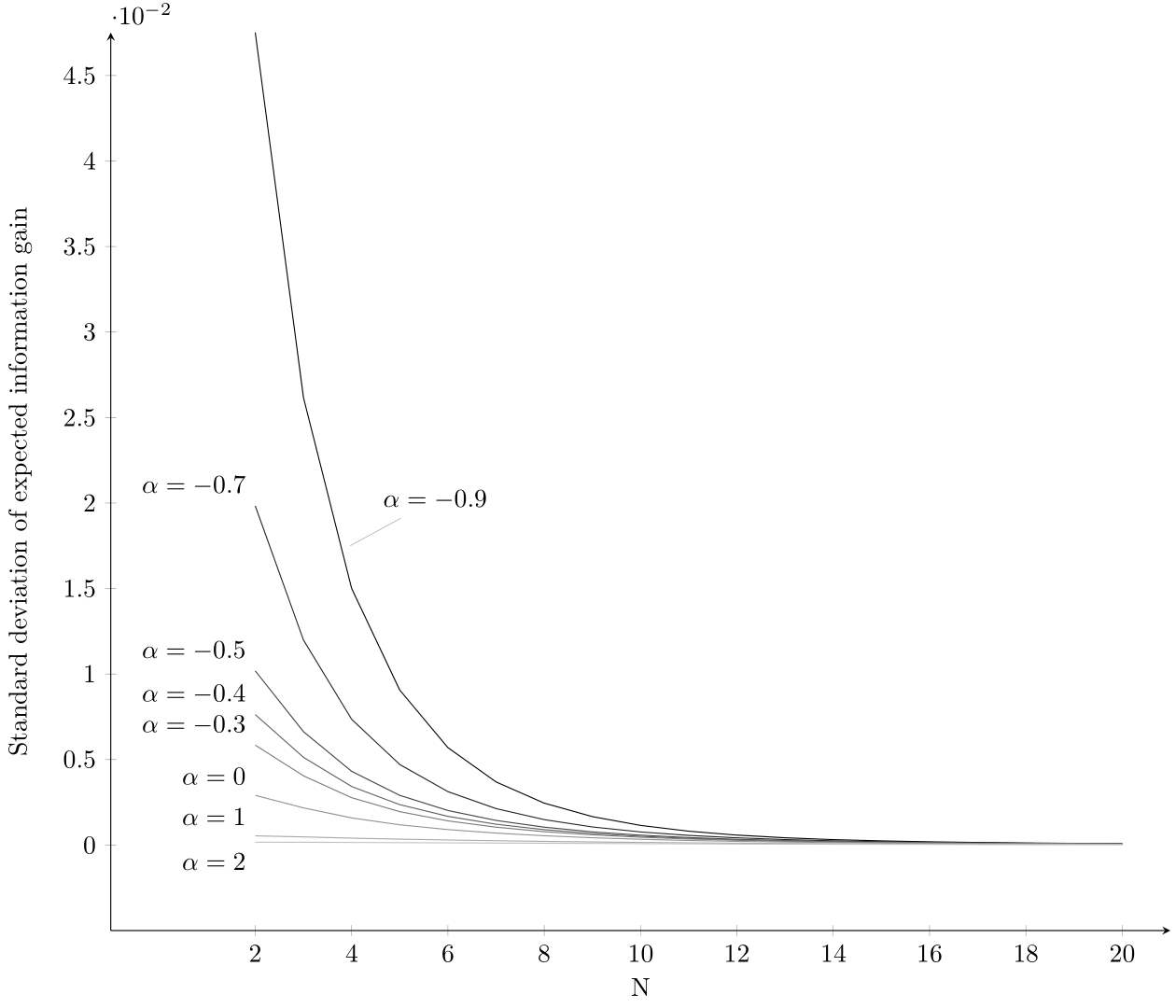}
		\caption{\label{expect_stddev} \textit{Robustness of Expected Information Gain.} The $y$-axis represents the standard deviation of the expected information gain over all possible $h_N$, while the $x$-axis represents the value of $N$. As $N$ increases, and even for relatively small values of $N$, the standard deviation tends toward zero for all priors.}
	\end{figure}

	As with the discussions of differential information gain and relative information gain, we are also interested in examining the dependence of expected information gain on $\alpha$ or $h_N$. However, such dependence appears to be weak, as illustrated in Figure \ref{expect_graph} and Figure \ref{expect_stddev}. Expected information gain demonstrates strong robustness concerning variations in $\alpha$ and $h_N$.
	
	The asymptotic expression of expected information gain is
	\begin{equation}
		\overline{I_{\text{diff}}} = \overline{I_{\text{rel}}} = \frac{1}{2N}
	\end{equation}
	
	\section{Comparison of Three Information Gain Measures, and the Information Increase Principle}
	\label{sec:comparisons}
	
	
	From an operational perspective, the information measures we have considered can be categorized into two types: differential information gain and relative information gain pertain to a measurement that has \emph{already been made}, while expected information gain pertains to a measurement that has \emph{yet to be conducted.}
	
	As regards to positivity, which is tied to the fundamental question of ``Will acquiring more data from measurements lead to a deeper understanding of the system?":~for relative information gain and expected information gain, the answer is affirmative; but, differential information gain is positive only under certain specific prior conditions.
	
	All three measures are functions of variables denoted as $N$, $\alpha$, and $h_N$, which characterize the size of the data sequences, the prior information, and the existing data sequence, respectively. How sensitive are these measures to these parameters, particularly for large values of $N$?
	As we have shown, differential information gain is heavily influenced by all three parameters. It becomes nearly independent of $h_N$ only when $\alpha = -0.5$. Relative information gain is not highly sensitive to the choice of priors. In the case of large $N$, relative information gain is affected by both $h_N$ and $N$, whereas expected information gain depends solely on $N$.

	\begin{table}[]
		\centering
		\label{comparison_measures}
		{\scriptsize
		\begin{tabular}{|p{2cm}|p{6.5cm}|p{6.5cm}|}
			\hline
			& Positivity
			& Robustness about~$T_N$ \\ \hline
			Diff Infor. Gain & Strictly positive when~$\alpha<\alpha_p$ where~$\alpha_p \approx -0.68$. Asymptotic positive when~$\alpha\le-0.5$. & Robustness exists only when~$\alpha=-0.5$ of beta distribution prior. \\ \hline
			Rel Infor. Gain & Strictly positive for all priors. & No significant differences of robustness among beta distribution priors. \\ \hline
			Exp. Infor. Gain & Strictly positive for all priors. & No significant differences of robustness among beta distribution priors. \\ \hline
		\end{tabular}}
		\caption{\textit{Comparison of Three Information Gain Measures}}
	\end{table}
	
	
	
	
	At first, one might have expected that the idea that \emph{more data from measurements lead to more knowledge about the system} would hold strictly, namely that the information gain from additional data would always be strictly positive. However, our perspective has been challenged by the observation of black swan events. In the extreme scenario where the first $N$ tosses all result in tails, and the $(N+1)$th toss yields a head, a negative information gain in this $(N+1)$th toss may be a more reasonable interpretation. To address this, we propose the Principle of Information Increase:
	\textit{In a series of binomial distribution data, the information gain from additional data should tend towards positivity in the asymptotic limit. However, in the extreme case where the first $N$ data points are identical and the data of the $(N+1)$th trial is opposite to the previous data, the information gain in this exceptional case should be negative.}
	
	Applying this criterion, the choice of using the differential information gain becomes more appropriate for measuring the extent of knowledge contributed by additional data. For the beta distribution prior, it should be constrained within the range of approximately $-0.68 \lesssim \alpha \le -0.5$. If we also consider the robustness of information gain under various given data scenarios, then the Jeffreys' binomial prior ($\alpha = -0.5$) emerges as the most favorable choice.
	
	
	\section{Related Work}
	\label{sec:related-work}
	
	\subsection{Information Increase Principle and Jeffreys' binomial prior}
	In \cite{Summhammer1994, Summhammer1999}, Summhammer introduces the idea that \emph{more measurements lead to more knowledge about a physical quantity}, and quantifies the level of knowledge regarding a quantity by assessing its uncertainty range after a series of repeated measurements.  Quantified in this manner, the notion can be summarized as: `The uncertainty range of a physical quantity should decrease as the number of measurements increases.' For a quantity~$\theta$, the uncertainty range~$\Delta \theta$ is a function of the number of measurements:
	\begin{equation}
		\label{Summhammer_uncer_increase}
		\Delta \theta  (N+1) < \Delta \theta (N)
	\end{equation}
	If this quantity is determined by the probability of a two-outcome measurement, such as the probability of obtaining heads ($p$) in a coin toss, then there exists a relationship between the uncertainty range of $\theta$ and that of $p$,
	\begin{equation}
		\Delta \theta = \left|\frac{\partial\theta}{\partial p}\right|\Delta p
	\end{equation}
	In large~$N$ approximation, $\Delta p = \sqrt{p(1-p)/N}$, so that 
	\begin{equation} \label{eqn:Summhammer-const}
		\Delta \theta  = \left|\frac{\partial\theta}{\partial p}\right| \sqrt{p(1-p)/N},
	\end{equation}
One intuitive way to ensure Eq.\eqref{Summhammer_uncer_increase} holds is by forcing $\Delta \theta$ to be purely a function of $N$. Observing the relationship between $\Delta \theta$ and $\Delta p$, the simplest solution would be to set $\Delta \theta = \frac{\text{const.}}{\sqrt{N}}$. Under this solution, the relationship between $p$ and $\theta$ takes the following form:
	\begin{equation}
		\left|\frac{\partial\theta}{\partial p}\right| \sqrt{p(1-p)} = \text{const.},
	\end{equation}
which yields Malus' law~$p(\theta) = \cos^2(m(\theta-\theta_0)/2)$ with~$m\in\numberfield{Z}$.

Summhammer does not employ information theory to quantify `knowledge about a physical quantity', but instead utilizes the statistical uncertainty associated with the quantity.  However, viewed from the Bayesian perspective, if we assume that the prior distribution of the physical quantity,~$\theta$, is uniform, the between~$\theta$ and~$p$ in Eq.~\eqref{eqn:Summhammer-const}  implies that the prior distribution of the probability follows Jeffreys' binomial prior,
	\begin{equation}
		\Pr(p|I) = \left|\frac{\partial\theta}{\partial p}\right| \Pr(\theta|I) = \frac{1}{\pi}\frac{1}{\sqrt{p(1-p)}}
	\end{equation}
Thus, in the large $N$ approximation, Summhammer's result can be interpreted to mean that the prior associated with the probability of a uniformly-distributed 
physical quantity must adhere to Jeffreys' binomial prior.  
	
	Goyal \cite{Goyal2005} introduces an \emph{asymptotic} Principle of Information Gain~(which differs from ours), which states that ``In $n$ interrogations of a $N$-outcome probabilistic source with an unknown probabilistic vector $\vec{P}$, the amount of Shannon-Jaynes information provided by the data about $\vec{P}$ remains independent of $\vec{P}$ for all $\vec{P}$ in the limit as $n\rightarrow\infty$.'' Goyal establishes the equivalence between this principle and Jeffreys' rule. Under his Principle of Information Gain, the Jeffreys' multinomial prior is then derived. In the case of a two-outcome probabilistic model, the Jeffreys' multinomial prior reduces to Jeffreys' binomial prior. Asymptotic analysis reveals that Shannon-Jaynes information is not only independent of the probability vector $\vec{P}$, but also monotonically increases with the number of interrogations. It is worth noting that Shannon-Jaynes information can be viewed as the accumulation of differential information gain. This asymptotic result aligns with our findings: under Jeffreys' binomial prior, the differential information gain is solely dependent on the number of measurements.
	
	\subsection{Other Information-theoretical motivations
	 of Jeffreys' binomial prior}
	
	Wootters \cite{Wotters2013} introduces a novel perspective on Jeffreys' binomial prior, where quantum measurement is employed as a communication channel. In this framework, 
	Alice aims to transmit a continuous variable, denoted as $\theta$, to Bob. Instead of directly sending $\theta$ to Bob, Alice transmits a set of identical coins to Bob, where the probability of getting heads, $p(\theta)$, in each toss is a function of $\theta$. Bob's objective is to maximize the information about $\theta$ that he can extract from a finite number of tosses. The measure of information used in this context is the mutual information between $\theta$ and the total number of heads, $n$, in $N$ tosses.
	\begin{equation}
		\label{Wooters_mutual_infor}
		I(n:\theta) = H(n) - H(n|\theta) = -\sum_{n=0}^{N}p(n)\ln P(n) - \left\langle-\sum_{n=0}^{N}p(n|p(\theta))\ln p(n|p(\theta)) \right\rangle
	\end{equation}

	However, the function $p(\theta)$ is unknown, and the optimization process begins with a set of discrete values, ${p_1, p_2, \ldots, p_L}$, rather than utilizing the continuous function $p(\theta)$. For each discrete value, $p_k$, there is an associated weight, $w_k$. The mutual information can be expressed as follows:
	\begin{equation}
		I(n:\theta) = -\sum_{n=0}^{N}p(n)\ln P(n) + \sum_{k=1}^{L}w_k\sum_{n=0}^{N}p(n|p_k)\ln p(n|p_k)
	\end{equation}
	In the large $N$ approximation, it is found that the weight $w$ takes on a specific form:
	\begin{equation}
		w(p) = \frac{1}{\pi\sqrt{p(1-p)}}
	\end{equation}
	which serves a role akin to the prior probability of $p$. Remarkably, this prior probability aligns with Jeffreys' binomial prior. A similar procedure can be extended to Jeffreys' multinomial prior distribution. Wootters' approach shares similarities with the concept of a reference prior, where the selected prior aims to maximize mutual information, which can be viewed as the expected information gain across all data. The outcome is consistent with the reference prior for multinomial data \cite{BergerBernardo1992}, thus revealing another informational interpretation of Jeffreys' prior.

	\section{Conclusion}
	\label{conclusion}
	
	In this paper, we delve into the concept of information gain for two-outcome quantum systems from an operational perspective. We introduce an informational postulate, the Principle of Information Increase, which serves as a criterion for selecting the appropriate measure to quantify the extent of information gained from measurements and the choice of prior. Our investigation reveals that the differential information gain emerges as the most physically meaningful measure when compared to another contender, the relative information gain.
	
	The Jeffreys' binomial prior exhibits notable characteristics within the realm of two-outcome quantum systems. Both Summhammer and our work demonstrate that under this prior, the intuitive notion that \emph{more data from measurements leads to more knowledge about the system} holds true, as confirmed by two distinct methods of quantifying knowledge. Additionally, Wootters shows that this prior enables the communication of maximal information, further highlighting its significance. We also find that Jeffreys' binomial prior displays robustness, although the origin of this robustness remains unexplained. It raises an intriguing question of whether this feature could be extended to multinomial distributions or other types of probabilistic systems. We speculate that there might be another layer of intuitive understanding related to the robustness of Jeffreys' prior.
	
	While this paper primarily focuses on the single-parameter beta distribution prior for binomial distributions, we anticipate similar results could manifest in multinomial distributions. However, it remains an open question how differential information gain behaves under more general types of priors and distributions.

	\bibliographystyle{plain}
	\bibliography{InforIncrea}
	
	\appendix
	\section{Derivation of Differential Information Gain}
	\label{appendix:derivation_dig}
	
	The posterior is determined by $T_N$ and prior. For the sake of simplicity we would set the prior belongs to the family of beta distributions:
	\begin{equation}
		\Pr(p|I)=\frac{p^{\alpha}(1-p)^{\alpha}}{B(\alpha+1,\alpha+1)}
	\end{equation}
	where $\alpha>-1, B(x,y)$ is the beta function.
	
	Given $N$, there are $2^N$ different $T_N$. However, we may not need to calculate all the $2^N$ sequences. Suppose every toss is independent, this happens in quantum mechanics, then this coin tossing model would become a binomial distribution. Let $h_N$ be the number of heads inside $T_N$, the posterior $\Pr(p|N,T_N,I)$ is equivalent to $\Pr(p|N,h_N,I)$ and likelihood will be
	\begin{equation}
		\Pr(h_N|N,p,I)={{N}\choose{h_N}} p^{h_N}(1-p)^{N-h_N}
	\end{equation}
	hence the posterior after $N$ tosses
	\begin{equation}
		\begin{aligned}
			\label{posterior}
			\Pr(p|N,h_N,I) &=\frac{\Pr(h_N|N,p,I)\Pr(p|I)}{\int \Pr(h_N|N,p,I)\Pr(p|I) dp}\\
			&=\frac{p^{h_N+\alpha}(1-p)^{N-h_N+\alpha}}{B(h_N+\alpha+1,N-h_N+\alpha+1)}
		\end{aligned}
	\end{equation}
	
	The information gain in the $(N+1)$th toss would be
	\begin{equation}
		\label{deltaH}
		I_{\text{diff}} = D_\text{KL}(\Pr(p|N+1,\{T_N,t_{N+1}\},I)||\Pr(p|I) ) - D_\text{KL}(\Pr(p|N,h_N,I)||\Pr(p|I) )
	\end{equation}
	
	$I_{\text{diff}}$ is determined by $h_N$, prior and the result of $(N+1)$th toss $t_{N+1}$. $t_{N+1}$ could be either \text{`Head'} or \text{`Tail'}, then posterior after $N+1$ tosses could be
	\begin{equation}
		\Pr(p|N+1,\{T_N,t_{N+1}=\text{`Head'}\text{`Head'}\},I)=\frac{p^{h_N+\alpha+1}(1-p)^{N-h_N+\alpha}}{B(h_N+\alpha+2,N-h_N+\alpha+1)}
	\end{equation}
	\begin{equation}
		\Pr(p|N+1,\{T_N,t_{N+1}=\text{`Tail'}\},I)=\frac{p^{h_N+\alpha}(1-p)^{N-h_N+\alpha+1}}{B(h_N+\alpha+1,N-h_N+\alpha+2)}
	\end{equation}
	
	{Taking~$t_{N+1} = \text{`Head'} $,the first term in (\ref{deltaH}) would become}
	\begin{equation}
		\begin{aligned}
			&D_\text{KL}(\Pr(p|N+1,\{T_N,t_{N+1}=\text{`Head'}\},I)||\Pr(p|I) )\\
			&=\int_0^1 \Pr(p|N+1,h_N+1,I)\ln\frac{\Pr(p|N+1,h_N+1,I)}{\Pr(p|I)}dp\\
			&=\int_0^1\frac{p^{h_N+\alpha+1}(1-p)^{N-h_N+\alpha}}{B(h_N+\alpha+2,N-h_N+\alpha+1)}\ln\frac{p^{h_N+1}(1-p)^{N-h_N}B(\alpha+1,\alpha+1)}{B(h_N+\alpha+2,N-h_N+\alpha+1)}dp\\
			&=\int_0^1\frac{p^{h_N+\alpha+1}(1-p)^{N-h_N+\alpha}}{B(h_N+\alpha+2,N-h_N+\alpha+1)} \left\{ \ln[p^{h_N+1}(1-p)^{N-h_N}]+\ln\frac{B(\alpha+1,\alpha+1)}{B(h_N+\alpha+2,N-h_N+\alpha+1)}\right\} dp\\
			&=\int_0^1\frac{p^{h_N+\alpha+1}(1-p)^{n-h_N+\alpha}}{B(h_N+\alpha+1,n-h_N+\alpha+1)} \ln[p^{h_N+1}(1-p)^{N-h_N}] dp+\ln\frac{B(\alpha+1,\alpha+1)}{B(h_N+\alpha+2,n-h_N+\alpha+1)}
		\end{aligned}
	\end{equation}
	
	By using the integral
	
	\begin{equation}
		\int_0^1 x^{a}(1-x)^{b}ln(x)dx=B(a+1,b+1)[\psi(a+1)-\psi(a+b+2)]
	\end{equation}
	where $\psi(x)$ is the digamma function\footnote{The digamma function can be defined in terms of gamma function: $\psi(x) = \frac{\Gamma'(x)}{\Gamma(x)} $.}, we can obtain the following result
	
	\begin{equation}
		\begin{aligned}
			D_\text{KL}(\Pr(p|N+1,\{T_N,t_{N+1}=\text{`Head'}\},I)||\Pr(p|I))=&(h_N+1)\psi(h_N+\alpha+2) + (N-h_N)\psi(N-h_N+\alpha+1) \\
			&-(N+1)\psi(N+2\alpha+3) \\
			&+\ln\frac{B(\alpha+1,\alpha+1)}{B(h_N+\alpha+2,n-h_N+\alpha+1)}
		\end{aligned}
	\end{equation}

	The second term in (\ref{deltaH}) would become
	\begin{equation}
		\begin{aligned}
			&D_\text{KL}(\Pr(p|N,h_N,I)||\Pr(p|I))\\
			&=\int_0^1 \Pr(p|N,h_N,I)\ln\frac{\Pr(p|N,h_N,I)}{\Pr(p|I)}dp\\
			&=\int_0^1\frac{p^{h_N+\alpha}(1-p)^{N-h_N+\alpha}}{B(h_N+\alpha+1,N-h_N+\alpha+1)}\ln\frac{p^{h_N}(1-p)^{N-h_N}B(\alpha+1,\alpha+1)}{B(h_N+\alpha+1,N-h_N+\alpha+1)}dp\\
			&=\int_0^1\frac{p^{h_N+\alpha}(1-p)^{N-h_N+\alpha}}{B(h_N+\alpha+1,N-h_N+\alpha+1)} \left\{ \ln[p^{h_N}(1-p)^{N-h_N}]+\ln\frac{B(\alpha+1,\alpha+1)}{B(h_N+\alpha+1,N-h_N+\alpha+1)}\right\} dp\\
			&=\int_0^1\frac{p^{h_N+\alpha}(1-p)^{n-h_N+\alpha}}{B(h_N+\alpha+1,n-h_N+\alpha+1)}\ln[p^{h_N}(1-p)^{N-h_N}]dp+\ln\frac{B(\alpha+1,\alpha+1)}{B(h_N+\alpha+1,n-h_N+\alpha+1)}\\
			&=h_N\psi(h_N+\alpha+1)+(N-h_N)\psi(N-h_N+\alpha+1)-N\psi(N+2\alpha+2) + \ln\frac{B(\alpha+1,\alpha+1)}{B(h_N+\alpha+1,n-h_N+\alpha+1)}
		\end{aligned}
	\end{equation}

	Now we obtain the final expression of (\ref{deltaH})
	
	\begin{equation}
		\begin{aligned}
			I_{\text{diff}}(t_{N+1}=\text{`Head'})=&~D_\text{KL}(\Pr(p|N+1,\{T_N,t_{N+1}=\text{`Head'}\},I)||\Pr(p|I) ) - D_\text{KL}(\Pr(p|N,h_N,I)||\Pr(p|I))\\
			=&~\psi(h_N+\alpha+2)-\psi(N+2\alpha+3)\\ 
			&+\frac{h_N}{h_N+\alpha+1}-\frac{N}{N+2\alpha+2}
			+\ln{\frac{N+2\alpha+2}{h_N+\alpha+1}}
		\end{aligned}
	\end{equation}

	Similarly we can obtain the~$I_{\text{diff}}$ when~$t_{N+1} = \text{`Tail'}$
	\begin{equation}
		\begin{aligned}
			I_{\text{diff}}(t_{N+1}=\text{`Tail'})=&~\psi(N-h_N+\alpha+2)-\psi(N-h_N+2\alpha+3)\\ 
			&~+\frac{N-h_N}{N-h_N+\alpha+1}-\frac{N}{N+2\alpha+2}
			+\ln{\frac{N+2\alpha+2}{N-h_N+\alpha+1}}
		\end{aligned}
	\end{equation}

	This suggests that for fixed~$N$ and~$\alpha$, $I_{\text{diff}}(t_{N+1}=\text{`Head'})$ and~$I_{\text{diff}}(t_{N+1}=\text{`Tail'})$ are symmetric since~$h_N$ is ranging from~$0$ to~$N$.
	
	\section{Derivation of Relative Information Gain}
	\label{appendix:derivation_rig}
	
	From Appendix A we know that the posterior after $N$ tosses is
	\begin{equation}
		\Pr(p|N,T_N,I)=\Pr(p|N,h_N,I)=\frac{p^{h_N+\alpha}(1-p)^{N-h_N+\alpha}}{B(h_N+\alpha+1,N-h_N+\alpha+1)}
	\end{equation}
	Therefore the posterior after $N+1$ tosses would be
	\begin{equation}
		\label{post-posterior}
		\Pr(p|N+1,T_{N+1},I)=\frac{\Pr(h_N,T_{N+1}|p,N+1,I) \Pr(p|I)}{\int_{0}^{1} \Pr(h_N,T_{N+1}|p,N+1,I) \Pr(p|I) dp}
	\end{equation}
	
	Depends on different results of $t_{N+1}$, the posterior after $N+1$ tosses would be
	\begin{align}
		&\Pr(p|N+1,\{T_N,t_{N+1}=\text{`Head'}\},I) = \frac{p^{h_N+\alpha+1} (1-p)^{N-h_N+\alpha}}{B(h_N+\alpha+2,N-h_N+\alpha+1)}\\
		&\Pr(p|N+1,\{T_N,t_{N+1}=\text{`Tail'}\},I) = \frac{p^{h_N+\alpha} (1-p)^{N-h_N+\alpha+1}}{B(h_N+\alpha+1,N-h_N+\alpha+2)}
	\end{align}
	
	And the corresponding relative information gain would be
	\begin{equation}
		\begin{aligned}
			&I_{\text{rel}}(t_{N+1}=\text{`Head'})\\
			&=D_\text{KL}(\Pr(p|N+1,\{T_N,t_{N+1}=\text{`Head'}\},I)||\Pr(p|N,h_N,I))\\
			&=\int_{0}^{1} \Pr(p|N+1,\{T_N,t_{N+1}=\text{`Head'}\},I) \ln{\frac{\Pr(p|N+1,\{T_N,t_{N+1}=\text{`Head'}\},I)}{\Pr(p|N,h_N,I)}} dp\\
			&=\int_{0}^{1} \frac{p^{h_N+\alpha+1} (1-p)^{N-h_N+\alpha}}{B(h_N+\alpha+2,N-h_N+\alpha+1)} \ln{ \frac{pB(h_N+\alpha+1,N-h_N+\alpha+1)}{B(h_N+\alpha+2,N-h_N+\alpha+1)}} dp\\
			&=\psi(h_N+\alpha+2)-\psi(N+2\alpha+3)+\ln{\frac{N+2\alpha+2}{h_N+\alpha+1}}
		\end{aligned}
	\end{equation}
	\begin{equation}
		I_{\text{rel}}(t_{N+1}=\text{`Tail'}) = \psi(N-h_N+\alpha+2)-\psi(N-h_N+2\alpha+3)+\ln{\frac{N+2\alpha+2}{N-h_N+\alpha+1}}
	\end{equation}
	
	\section{Equivalence of Expected Differential Information Gain and Expected Relative Information Gain}
	\label{appendix:equivalence_exp}
	
	From~~(\ref{ave_p}), we know
	\begin{equation}
		\langle p \rangle = \int_{0}^{1} p~\Pr(p|N,h_N,I)~dp
	\end{equation}
	
	If the~$(N+1)$th toss is~\text{`Head'}, then the posterior after $N+1$ tosses can be written as
	\begin{equation}
		\Pr(p|N+1,\{T_N,t_{N+1}=\text{`Head'}\},I) = \frac{p~\Pr(p|N,h_N,I)}{\int_{0}^{1} p~\Pr(p|N,h_N,I)~dp} = \frac{p~\Pr(p|N,h_N,I)}{\langle p \rangle}
	\end{equation}
	
	Then we can rewrite $I_{\text{diff}}$ as
	\begin{equation}
		\begin{aligned}
			I_{\text{diff}}(t_{N+1}=\text{`Head'}) =& D_\text{KL}(\Pr(p|N+1,\{T_N,t_{N+1}=\text{`Head'}\},I)||\Pr(p|I))-D_\text{KL}(\Pr(p|N,h_N,I)||\Pr(P|I))\\
			=& \int_{0}^{1} \frac{p~\Pr(p|N,h_N,I)}{\langle p \rangle} \ln{\frac{p~\Pr(p|N,h_N,I)}{\langle p \rangle~\Pr(p|I)}}~dp - \int_{0}^{1} \Pr(p|N,h_N,I) \ln{\frac{\Pr(p|N,h_N,I)}{\Pr(p|I)}}~dp\\
			=& \int_{0}^{1} \frac{p~\Pr(p|N,h_N,I)}{\langle p \rangle} \left(\ln{\frac{\Pr(p|N,h_N,I)}{\Pr(p|I)}}+\ln{\frac{p}{\langle p \rangle}}\right)~dp \\
			&- \int_{0}^{1} \Pr(p|N,h_N,I) \ln{\frac{\Pr(p|N,h_N,I)}{\Pr(p|I)}}~dp
		\end{aligned}
	\end{equation}
	
	Similarly if~$t_{N+1} = \text{`Tail'}$,
	\begin{equation}
		\begin{aligned}
			I_{\text{diff}}(t_{N+1}=\text{`Tail'}) =& D_\text{KL}(\Pr(p|N+1,\{T_N,t_{N+1}=\text{`Tail'}\},I)||\Pr(p|I))-D_\text{KL}(\Pr(p|N,h_N,I)||\Pr(p|I))\\
			=& \int_{0}^{1} \frac{(1-p)~\Pr(p|N,h_N,I)}{\langle 1-p \rangle} \left(\ln{\frac{\Pr(p|N,h_N,I)}{\Pr(p|I)}}+\ln{\frac{1-p}{\langle 1-p \rangle}}\right)~dp \\
			&- \int_{0}^{1} \Pr(p|N,h_N,I) \ln{\frac{\Pr(p|N,h_N,I)}{\Pr(p|I)}}~dp
		\end{aligned}
	\end{equation}
	Then the expected differential information gain would be
	\begin{equation}
		\label{expect_dig_formula}
		\begin{aligned}
			\overline{I_{\text{diff}}} &= \langle p \rangle \times I_{\text{diff}}(t_{N+1}=\text{`Head'}) + \langle 1-p \rangle \times I_{\text{diff}}(t_{N+1}=\text{`Tail'})\\
			&= \int_{0}^{1} p~\Pr(p|N,h_N,I) \left(\ln{\frac{\Pr(p|N,h_N,I)}{\Pr(p|I)}}+\ln{\frac{p}{\langle p \rangle}}\right)~dp \\
			&\quad+ \int_{0}^{1} (1-p)~\Pr(p|N,h_N,I) \left(\ln{\frac{\Pr(p|N,h_N,I)}{\Pr(p|I)}} +\ln{\frac{1-p}{\langle 1-p \rangle}}\right)~dp \\
			&\quad- \int_{0}^{1} \Pr(p|N,h_N,I) \ln{\frac{\Pr(p|N,h_N,I)}{\Pr(p|I)}}~dp \\
			&= \int_{0}^{1} \Pr(p|N,h_N,I) \left(p~\ln{\frac{p}{\langle p \rangle}} + (1-p)~\ln{\frac{1-p}{\langle 1-p \rangle}}\right)~dp
		\end{aligned}
	\end{equation}
	
	Similarly, $I_{\text{rel}}$ can be written as
	\begin{equation}
		\begin{aligned}
			I_{\text{rel}}(t_{N+1}=\text{`Head'}) &= D_\text{KL}(\Pr(p|N+1,\{T_N,t_{N+1}=\text{`Head'}\},I)||\Pr(p|N,h_N,I)) \\
			&= \int_{0}^{1} \frac{p~\Pr(p|N,h_N,I)}{\langle p \rangle} \ln{\frac{p}{\langle p \rangle}}~dp
		\end{aligned}
	\end{equation}
	
	\begin{equation}
		\begin{aligned}
			I_{\text{rel}}(t_{N+1}=\text{`Tail'}) &= D_\text{KL}(\Pr(p|N+1,\{T_N,t_{N+1}=\text{`Tail'}\},I)||\Pr(p|N,h_N,I)) \\
			&= \int_{0}^{1} \frac{(1-p)~\Pr(p|N,h_N,I)}{\langle 1-p \rangle} \ln{\frac{1-p}{\langle 1-p \rangle}}~dp
		\end{aligned}
	\end{equation}
	Then the expected relative information gain would be
	\begin{equation}
		\label{expect_rig_formula}
		\begin{aligned}
			\overline{I_{\text{rel}}} &= \langle p \rangle \times I_{\text{rel}}(t_{N+1}=\text{`Head'}) + \langle 1-p \rangle \times I_{\text{rel}}(t_{N+1}=\text{`Tail'})\\
			&= \int_{0}^{1} p~\Pr(p|N,h_N,I)~\ln{\frac{p}{\langle p \rangle}}~dp + \int_{0}^{1} (1-p)~\Pr(p|N,h_N,I)~\ln{\frac{1-p}{\langle 1-p \rangle}}~dp \\
			&= \int_{0}^{1} \Pr(p|N,h_N,I) \left(p~\ln{\frac{p}{\langle p \rangle}} + (1-p)~\ln{\frac{1-p}{\langle 1-p \rangle}}\right)~dp
		\end{aligned}
	\end{equation}
	
	From (\ref{expect_dig_formula}) and (\ref{expect_rig_formula}) we may see that in this two-outcome model, the expected differential information gain $\overline{I_{\text{diff}}}$ and expected relative information gain $\overline{I_{\text{rel}}}$ are equal under all kinds of priors.

\end{document}